\documentclass[lettersize,journal]{IEEEtran}
\usepackage{amsmath,amsfonts}
\usepackage{algorithmic}
\usepackage{algorithm}
\usepackage{array}
\usepackage[caption=false,font=normalsize,labelfont=sf,textfont=sf]{subfig}
\usepackage{textcomp}
\usepackage{stfloats}
\usepackage{url}
\usepackage{verbatim}
\usepackage{graphicx}
\usepackage{cite}
\usepackage{hyperref}
\begin{document}

\title{Sensorless Adaptive Vibration Suppression in Two-Mass Systems via Joint Estimation of Controller Parameters and System States}

\author{A. Glushchenko,~\IEEEmembership{Member,~IEEE,}
        K. Lastochkin
\thanks{Financial support is in part provided by the Grants Council of the President of the Russian Federation (MD-1787.2022.4)}
\thanks{Anton Glushchenko and Konstantin Lastochkin are with V.A. Trapeznikov Institute of Control Sciences of Russian Academy of Sciences, Moscow 117997, Russia
(e-mail: aiglush@ipu.ru; lastconst@ipu.ru).}}



\maketitle

\begin{abstract}
The scope of this study is to develop a novel sensorless adaptive vibration suppression controller for two-mass systems with joint estimation of states and controller parameters. Unlike existing solutions, we simultaneously: \linebreak ({\it i}) propose an analytically proved, unified and singularity-issue-free scheme of parameters adjustment of a control law with additional feedbacks that ensures convergence of such parameters to their true values under extremely weak regressor finite excitation (FE) requirement, ({\it ii}) derive an adaptive observer of a two-mass electromechanical system physical states with guarantee of their convergence to the ground truth values under clear FE condition, ({\it iii}) rigorously prove the exponential stability of the obtained closed-loop system of adaptive vibration suppression for two-mass systems that includes the above-mentioned adaptive observer and adaptive controller. These approaches are grounded on the recently proposed method of parameters identification for one class of nonlinearly parameterized regression equation and thoroughly investigated dynamic regression extension and mixing procedure (DREM). The obtained theoretical results are confirmed via numerical experiments. 
\end{abstract}


\section{Introduction}
\IEEEPARstart{A}{n} electromechanical system composed of a power converter, an electrical drive, and a mechanical gearbox always has elastic properties, which are caused by the transfer of torque from the motor shaft to the load through elastic mechanical joints (elastic couplings, cable-pulley systems, drivebelts, long shafts, conveyor belts, etc.) \cite{b1}. Relevant industrial examples \cite{b2,b3,b4} of such systems are conveyers, rolling-mills, throttle drives, wind-mill turbines, paper machines, servo drives, robot arm drives and many others. 

A distinctive feature of two-mass systems is elastic vibrations of the mechanical system, which cause a) excessive wear of mechanics and electrics; b) lower accuracy of the reference tracking by the load and, as a consequence, reduction of the plant performance and the final product quality.

The conventional cascade control structure for two-mass systems is based on PI motor speed and current controllers and, as was thoroughly discussed in \cite{b5,b6}, not able to ensure the required damping factor and desired pole location. In order to improve the control quality, various approaches \cite{b6,b7,b8,b9,b10,b11} have been developed recently. The most advanced techniques, which provide better transient performance, are based on special control structures with additional feedbacks from states such as load speed, torsional/load torque and their combinations. In \cite{b7} an additional feedback from the elastic (shaft) torque derivative was introduced. In \cite{b8}, instead of this derivative, the elastic (shaft) torque was used to cope with the negative effect of the measurement noise. In \cite{b9} an extra feedback from the difference between the speed of the motor and the load was utilized. In \cite{b8} and \cite{b10} another feedback from the load speed was applied. The exhaustive overview of most of existing controllers of two-mass systems with additional feedbacks is presented in \cite{b6,b10,b11}. In contrast to conventional purely cascade control design, such approaches provide desired pole location and allows one to regulate a damping factor in a closed loop.

However, as far as a typical benchmark two-mass control problem is concerned, at least two drawbacks of the above-mentioned solutions are worth mentioning \cite{b12}. It is assumed that:
\begin{enumerate}
    \item[\textbf{D1}] the systems parameters are {\it a priori} known (the controller coefficients depend on the value of these parameters);
    \item[\textbf{D2}] the state variables required for the additional feedbacks are directly measurable (it is impracticable due to technical difficulties, excessive cost, and reduction of the system reliability).
\end{enumerate}

So, in order to suppress vibration in two-mass systems successfully, it is necessary to overcome these drawbacks, and hence jointly obtain states and parameters estimations in on-line mode \cite{b12}. Using special recurrent algorithms such as extended state-parametric observers, it is necessary to obtain estimates of both states and system parameters. State ones are used to introduce additional feedbacks, and parameter ones are utilized to calculate feedbacks coefficients.

Having analyzed the control literature, it is concluded that the Model Reference Adaptive Systems framework \cite{b13} is the most appropriate one to provide a mathematically sound way to deal with the problem under consideration. Unfortunately, conventional adaptive state observers proposed by K. Narendra \cite{b14} and G. Kreisselmeier \cite{b15} in the middle of 70$^{\rm th}$ are capable of system state and parameters reconstruction only in case when such system is represented in a special observer canonical form. The problem is that, instead of \underline{physical} states (e.g. load speed, load and torsional torque) and parameters (e.g. motor ones, load inertia and stiffness coefficient), only \underline{virtual} states (linear combinations of all physical states) and parameters (e.g. parameters of both numerator and denominator of a transfer function from motor torque to motor speed) are reconstructed/identified in such case. That is a reason why, to the best of authors’ knowledge, there are no successful applications of the conventional adaptive observers \cite{b13,b14,b15} to solve \textbf{D2} in the existing literature \cite{b12,b16,b17,b18,b19,b21,b22,b23}. 

An extended or nonlinear Kalman filter \cite{b12,b16,b17,b18,b19} is free from the above-mentioned drawbacks of adaptive observers and allows one to reconstruct the system states, the external perturbation, the unknown physical parameters, and take into account the effect of backlashes and other nonlinearities. However, the disadvantages of its application \cite{b12,b16,b17,b18,b19} are as follows: ({\it i}) the covariance matrices are required to be chosen correctly from some {\it a priori} assumptions, ({\it ii}) the exponential convergence of the estimates to their true values is ensured only if the restrictive condition of the regressor persistent excitation \cite[Appendix]{b20} is met, which is hard to be satisfied, especially in the extended parameter-state-disturbance spaces, ({\it iii}) singularity may occur as the division operation is used for direct recalculation of the system parameters into the control law ones (see \cite[Fig.2]{b16}, \cite[Fig.1]{b18}, \cite[Fig.2]{b19}, and also the equations to obtain the parameters of additional feedback from \cite[Section 3]{b12}).

As the classic adaptive observers are not applicable \cite{b13,b14,b15} and the Kalman-filter-based techniques \cite{b12,b16,b17,b18,b19} have the above-mentioned disadvantages, the intelligent approaches have also attracted much attention. In \cite{b21,b22,b23} it was proposed to reconstruct unmeasured states of a two-mass electromechanical system using observers based on artificial neural networks or fuzzy logic. Their effectiveness was demonstrated experimentally, whereas the theoretical proof of stability and convergence is questionable. So, the disadvantage of the solutions \cite{b21,b22,b23} is that a mathematically sound analysis of both the conditions, under which the obtained estimates converge to their true values, and the stability of the designed closed-loop system is unclear \cite{b24}.

In general, the properties of the analyzed observers from \cite{b12,b16,b17,b18,b19,b21,b22,b23} and the indirect adaptive control systems based on them have been demonstrated in experimental setup, but have only sophisticated proofs without {\it bona fide} analytical analyses (Lyapunov like, for example). Many problems are intentionally omitted from the discussion, e.g. singularity issue due to recalculation of system parameters into control law ones, excitation conditions requirements for convergence, stability conditions etc. Therefore, the motivation of this study is to develop a novel adaptive vibration suppression controller for two-mass systems with joint estimation of states and controller parameters and mathematically sound proofs of closed-loop stability. In comparison with \cite{b12,b16,b17,b18,b19,b21,b22,b23}, the contribution of this study is summarized as follows:
\begin{enumerate}
    \item[\textbf{C1}] a unified and singularity-issue-free scheme is proposed to adjust the parameters of control laws with additional feedbacks from \cite{b6}, which ensures the convergence of the estimates (parameters of both the motor speed PI controller and additional feedbacks) to their true values if the regressor is finitely exciting. This contribution is addressed to solve \textbf{D1} and, in comparison with the existing approaches \cite{b12,b16,b17,b18,b19,b21,b22,b23}, the proposed one is free from singularity issue, has strictly weaker convergence conditions and is strongly analytically proved.
    \item[\textbf{C2}] an adaptive observer of the physical states of two-mass electromechanical systems is derived, which guarantees exponential convergence of the estimates of the physical states (load speed and torque, torsional torque) to their true values if the regressor finite excitation condition is met. This contribution is addressed to solve \textbf{D2} and, in comparison with the existing approaches \cite{b12,b16,b17,b18,b19,b21,b22,b23}, the proposed one has clear convergence conditions and is proved in a mathematically sound way.
    \item[\textbf{C3}] the exponential stability of the obtained closed-loop system of adaptive vibration suppression for two-mass systems (adaptive observer (\textbf{C1}) + adaptive controller (\textbf{C2}) applied to two-mass system) is rigorously proved.
\end{enumerate}

The requirement of the regressor finite excitation, which is necessary to ensure convergence of estimates of both unmeasured physical states and unknown controller parameters, is weak enough for many practical applications and equivalent to the identifiability of the system parameters over a finite time interval. We use modern methods of continuous system identification and adaptive observer design to achieve stated goal and provide \textbf{C1}-\textbf{C3} contributions. Particularly, a recently developed approach \cite{b25} is applied to estimate physical states of linear time-invariant systems, and a method \cite{b26} of parameters identification for one class of nonlinearly parameterized regression equation is used to obtain controller parameters. They are based on the thoroughly investigated dynamic regression extension and mixing procedure \cite{b27}. 

The proposed adaptive system for two-mass system vibration suppression structurally replicates the known solutions (see \cite[Fig.2]{b16}, \cite[Fig.1]{b18}, \cite[Fig.2]{b19} and Fig.1 in this paper) and consists of three blocks corresponding to the contributions \textbf{C1}, \textbf{C2} and \textbf{C3}. As for the first one, the problem of physical states observation is reduced to the one of parameter identification (it follows from the convergence of some parameters identification error that the states observation error also converges to zero). Considering the second block, on the basis of measurable signals, a special parameterization scheme for the regression equation with respect to the parameters of the control laws from \cite{b6} is proposed. In the third block, the identification laws are proposed that guarantee the convergence of errors of the system states observation and the control law parameters identification to zero if the measurable signals are finitely exciting.

The main result of the study is an adaptive system to suppress vibrations of a two-mass elastic electromechanical system, which does not require the system physical states to be measurable and the system parameters to be known. The result is obtained on the basis of the method to estimate the linear systems physical states \cite{b25} and the approach \cite{b26} to identify the parameters of nonlinear regression equations with overparameterization.

The remainder of the paper is organized as follows. Section II provides a rigorous problem statement. In Section III a new adaptive observer of the two-mass systems physical states and a unified scheme to adjust the parameters of the control laws with additional feedbacks from \cite{b6} are elucidated. Section IV presents the results of numerical experiments. The paper is wrapped up with conclusions in Section V.

\textbf{Disclaimer}. For a complete and detailed understanding of some derivations, it is advantageous, although not essential, to first study the dynamic regressor extension and mixing procedure \cite{b27}, the method to design external deterministic perturbation observers \cite{b28} and adaptive state observers for linear systems with overparameterization \cite{b25}, and the method of exponentially stable adaptive control \cite{b29}.

\textbf{Notation and Definitions.} The following notation is used throughout the paper. $\mathbb{R}^n$ and $\mathbb{R}^{n \times m}$ denote the sets of $n$-dimensional real vectors and $n \times m$-dimensional real matrices, respectively, $\mathbb{R}_{^ > }$ is a set of positive real numbers, $|.|$ represents the absolute value, $\|.\|$ denotes Euclidean norm of a vector, $I_n$ is an identity square $n \times n$ matrix, $0_{n \times m}$ is nullity $n \times m$-dimensional matrix. $\lambda_{\min }\left( . \right)$ and $\lambda_{\max }\left( . \right)$ are the matrix minimum and maximum eigenvalues, respectively, $vec\left( . \right)$ stands for the matrix vectorization, $\otimes$ is the Kronecker product. ${\rm{det}}\{.\}$ stands for a matrix determinant, ${\rm{adj}}\{.\}$ is an adjoint matrix, $L_{\infty}$ is the space of all essentially bounded functions. We also use the fact that for all (possibly singular) ${n \times n}$ matrices $M$ the following holds: ${\rm{adj}} \{M\} M = {\rm{det}} \{M\}I_{n \times n}$.

\textbf{Definition 1.} \emph{A mapping ${\cal F}\left( x \right){\rm{:\;}}{\mathbb{R}^{{n_x}}} \!\to\! {\mathbb{R}^{{n_{\cal F}} \times {m_{\cal F}}}}$ is heterogenous of degree ${\ell _{\cal F}} \!\ge\! 1$ if there exist  ${\Pi _\mathcal{F}}\!\left( {\omega \left( t \right)} \right) \!\in\! {\mathbb{R}^{{n_\mathcal{F}} \times {n_\mathcal{F}}}}{\text{, }}{\Xi _\mathcal{F}}\left( {\omega \left( t \right)} \right) = {\overline \Xi _\mathcal{F}}\left( {\omega \left( t \right)} \right)\omega \left( t \right) \in {\mathbb{R}^{{\Delta _\mathcal{F}} \times {n_x}}}$, and a mapping ${{\cal T}_{\cal F}}\left( {{\Xi _{\cal F}}\left( {\omega \left( t \right)} \right)x} \right){\rm{:\;}}{\mathbb{R}^{{\Delta _{\cal F}}}} \to {\mathbb{R}^{{n_{\cal F}} \times {m_{\cal F}}}}$ such that for all $\omega \left( t \right) \in \mathbb{R}$ and $x \in {\mathbb{R}^{{n_x}}}$ the following conditions hold:}
\begin{equation}\label{eq1}
\begin{gathered}
{\Pi _{\cal F}}\left( {\omega \left( t \right)} \right){\cal F}\left( x \right) = {{\cal T}_{\cal F}}\left( {{\Xi _{\cal F}}\left( {\omega \left( t \right)} \right)x} \right){\rm{,\;}}\\
{\rm{det}}\left\{ {{\Pi _{\cal F}}\left( {\omega \left( t \right)} \right)} \right\} \ge {\omega ^{{\ell _{_{\cal F}}}}}\left( t \right){\rm{,}}\\
{\Xi _{\cal F}}_{ij}\left( {\omega \left( t \right)} \right) = {c_{ij}}{\omega ^\ell }\left( t \right){\rm{,\;}}{c_{ij}} \in \left\{ {0,{\rm{ 1}}} \right\}{\rm{,\;}}\ell  > 0.
\end{gathered}    
\end{equation}

For example, the mapping ${\cal F}\left( x \right) = {\rm{col}}\left\{ {{x_1}{x_2}{\rm{,\;}}{x_1}} \right\}$ with ${\Pi _{\cal F}}\left( \omega  \right) = {\rm{diag}}\left\{ {{\omega ^2}{\rm{,\;}}\omega} \right\}{\rm{,\;}}{\Xi _{\cal F}}\left( \omega  \right) = {\rm{diag}}\left\{ {\omega {\rm{,\;}}\omega } \right\}$ is heterogenous of degree ${\ell _{\cal F}} = 3.$

\textbf{Definition 2.} \emph{A regressor $\varphi \left( t \right) \in {\mathbb{R}^n}$ is finitely exciting $\varphi \left( t \right) \in {\rm{FE}}$ over the time range $\left[ {t_r^ + {\rm{;\;}}{t_e}} \right]$ if there exists $t_r^ +  \ge 0$, ${t_e} > t_r^ + $ and $\alpha $ such that the following inequality holds:}
\begin{equation}\label{eq2}
\int\limits_{t_r^ + }^{{t_e}} {\varphi \left( \tau  \right){\varphi ^{\rm{T}}}\left( \tau  \right)d} \tau  \ge \alpha I_n{\rm{,}}    
\end{equation}
\emph{where $\alpha > 0$ is the excitation level.}

\textbf{Corollary 1.} \emph{For any matrix $D > 0$, controllable pair $\left( {A{\rm{,\;}}B} \right)$ with $B \in {\mathbb{R}^{n \times m}}$ and a Hurwitz matrix $A \in {\mathbb{R}^{n \times n}}$ there exist matrices $P = {P^{\rm{T}}} > 0$, $Q \in {\mathbb{R}^{n \times m}},\;K \in {\mathbb{R}^{m \times m}}$ and a scalar $\mu  > 0$ such that:}
\begin{equation}\label{eq3}
\begin{gathered}
{A^{\rm{T}}}P + PA =  - Q{Q^{\rm{T}}} - \mu P{\rm{,\;}}PB = QK{\rm{,\;}}\\
{K^{\rm{T}}}K = D + {D^{\rm{T}}}.
\end{gathered}
\end{equation}

\section{Problem statement}
A classic model of two-mass electromechanical system with neglected inner torque loop \cite{b6} is considered:
\begin{equation}\label{eq4}
\begin{gathered}
  {{\dot x}_p}\left( t \right) = A\left( \theta  \right){x_p}\left( t \right) + B\left( \theta  \right)u\left( t \right) + D\left( \theta  \right)\delta \left( t \right){\text{,}} \hfill \\
  y\left( t \right) = {C^{\text{T}}}{x_p}\left( t \right){\text{, }}x\left( {{t_0}} \right) = {x_0}{\text{,}} \hfill \\ 
\end{gathered}
\end{equation}
where
\begin{equation}\label{eq5}
\begin{gathered}
A\left( \theta  \right){\rm{:}} = {\begin{bmatrix}
0&0&{ - \theta _1^{ - 1}}\\
0&0&{\theta _2^{ - 1}}\\
{\theta _3^{ - 1}}&{ - \theta _3^{ - 1}}&0
\end{bmatrix}}{\rm{,\;}}B\left( \theta  \right){\rm{:}} = {\begin{bmatrix}
{\theta _1^{ - 1}}\\
0\\
0
\end{bmatrix}}{\rm{,\;}}\\
{C^{\rm{T}}} = {\begin{bmatrix}
1&0&0
\end{bmatrix}}{\rm{,\;}} D\left( \theta  \right){\rm{:}} = {\begin{bmatrix}
0&{ - \theta _2^{ - 1}}&0
\end{bmatrix}}^{\rm T}{\rm{,}}
\end{gathered}
\end{equation}
${x_{1p}}\left( t \right)$ and ${x_{2p}}\left( t \right)$ are the motor and load speeds, $u\left( t \right){\rm{,\;}}{x_{3p}}\left( t \right){\rm{,\;}}\delta \left( t \right)$ stand for the motor, torsional and load torques, respectively, $\theta  \in \mathbb{R}_{^ > }^3$ denotes a vector of unknown parameters including the mechanical time constant of motor ${\theta _1}$, load ${\theta _2}$ and elastic joint ${\theta _3}$. Only motor speed ${x_{1p}}\left( t \right)$ and torque $u\left( t \right)$ are measured with the help of sensors. The pairs $\left( {A\left( \theta  \right){\rm{,\;}}B\left( \theta  \right)} \right)$ and $\left( {{C^{\rm{T}}}{\rm{,\;}}A\left( \theta  \right)} \right)$ are completely observable and controllable. A class of considered load torque $\delta \left( t \right)$ signals is described in the following assumption.

\textbf{Assumption 1.} \emph{A disturbance $\delta \left( t \right)$ is a bounded and continuous signal formed by a time-invariant exosystem:}
\begin{equation}\label{eq6}
\begin{gathered}
{{\dot x}_\delta }\left( t \right) = {{\cal A}_\delta }{x_\delta }\left( t \right){\rm{,\;}}{x_\delta }\left( {{t_0}} \right) = {x_{\delta 0}}{\rm{,}} \hfill\\
\delta \left( t \right) = h_\delta ^{\rm{T}}{x_\delta }\left( t \right){\rm{,}}\hfill 
\end{gathered}
\end{equation}
\emph{where ${x_\delta }\left( t \right) \in {\mathbb{R}^{{n_\delta }}}$ are exosystem states with unknown initial conditions ${x_{\delta 0}}\left( {{t_0}} \right)$, ${h_\delta } \in {\mathbb{R}^{{n_\delta }}}{\rm{,\;}}{{\cal A}_\delta } \in {\mathbb{R}^{{n_\delta } \times {n_\delta }}}$ stand for known vector and matrix such that the pair $\left( {h_\delta ^{\rm{T}}{\rm{,\;}}{{\cal A}_\delta }} \right)$ is observable.}

As far as practical scenarios are concerned, a baseline control law for the system \eqref{eq4} is a PI-controller presented as\footnote{Regardless of the equation (8) and the structural diagram in Fig. 3 from \cite{b6}, a PI-controller \eqref{eq7} has been used to conduct the numerical experiments in \cite{b6} (e.g., Fig. 5 and Fig. 6).}:
\begin{equation}\label{eq7}
\begin{gathered}
{u_{bl}}\left( t \right) = {K_{{\rm{PI}}}}\left( \theta  \right){e_y}\left( t \right){\rm{,}}\\
{K_{{\rm{PI}}}}\left( \theta  \right) = {{\begin{bmatrix}
{{K_{\rm{P}}}\left( \theta  \right)}\\
{{K_{\rm{I}}}\left( \theta  \right)}
\end{bmatrix}}^{\rm{T}}}{\rm{,\;}}{e_y}\left( t \right){\rm{ = }}{\begin{bmatrix}
{ - y\left( t \right)}\\
{{e_I}\left( t \right)}
\end{bmatrix}}{\rm{,}}
\end{gathered}
\end{equation}
where ${e_I}\left( t \right) = \int\limits_{{t_0}}^t {{e_P}\left( \tau  \right)d\tau } $ is a reference $r\left( t \right) \in \mathbb{R}$ tracking integral error, ${e_P}\left( t \right) = r\left( t \right) - y\left( t \right)$ denotes reference tracking proportional error.

Using \eqref{eq7}, the considered model \eqref{eq4} is augmented with the integral error ${e_I}\left( t \right)$:
\begin{equation}\label{eq8}
\begin{gathered}
\dot x\left( t \right) = {\cal A}\left( \theta  \right)x\left( t \right) + {\cal B}\left( \theta  \right)u\left( t \right) + {\cal D}\left( \theta  \right)\delta \left( t \right){\rm{,}}\hfill\\
y\left( t \right) = {{\cal C}^{\rm{T}}}x\left( t \right){\rm{,}}\hfill
\end{gathered}
\end{equation}
where
\begin{equation}\label{eq9}
\begin{gathered}
x\left( t \right) = {\begin{bmatrix}
{{e_I}\left( t \right)}\\
{{x_p}\left( t \right)}
\end{bmatrix}}{\rm{,\;}}{\cal A}\left( \theta  \right) = {\begin{bmatrix}
0&{ - {C^{\rm{T}}}}\\
{{0_3}}&{A\left( \theta  \right)}
\end{bmatrix}}{\rm{,\;}}\\
{\cal B}\left( \theta  \right) = {\begin{bmatrix}
0\\
{B\left( \theta  \right)}
\end{bmatrix}}{\rm{,\;}}{\cal D}\left( \theta  \right) = {\begin{bmatrix}
0\\
{D\left( \theta  \right)}
\end{bmatrix}}{\rm{,\;}}\\
{{\cal C}^{\rm{T}}} = {\begin{bmatrix}
0&1&0&0
\end{bmatrix}}.
\end{gathered}
\end{equation}

The following assumption is adopted for $u\left( t \right)$.

\textbf{Assumption 2.} \emph{There exists a factorized full-information-based control law}
\begin{equation}\label{eq10}
\begin{gathered}
{u^*}\left( t \right) = {\kappa ^{\rm{T}}}\left( \theta  \right)x\left( t \right) = {u_{bl}}\left( t \right) + {K_{x} }\left( \theta  \right){x_p}\left( t \right) = \;\;\;\;\;\;\;\;\;\\\hfill
 = {K_{{\rm{PI}}}}\left( \theta  \right){e_y}\left( t \right) + {K_{x} }\left( \theta  \right){x_p}\left( t \right){\rm{,}}
\end{gathered}
\end{equation}
\emph{such that ${A_{ref}} = {\cal A}\left( \theta  \right) + {\cal B}\left( \theta  \right){\kappa ^{\rm{T}}}\left( \theta  \right)$ is a Hurwitz matrix, and the notation ${K_{x} }\left( \theta  \right) = {{\begin{bmatrix}
0&{{K_{1x}}\left( \theta  \right)}&{{K_{2x}}\left( \theta  \right)}
\end{bmatrix}}^{\rm{T}}}$ is introduced.}

The full-information-based control law ${u^*}\left( t \right)$ includes a baseline PI-controller \eqref{eq7} and additional feedbacks from the states ${x_p}\left( t \right)$, which are formed in accordance with \cite{b6} to improve transient quality, e.g. to suppress vibration. The choice of a particular additional feedback from the variety of existing ones \cite{b6} determines the desired quality indices of the closed-loop control system (damping, overshoot, etc.). Note that the authors of \cite{b6} propose to apply rather exotic control laws of the form ${u^*}\left( t \right) = v\left( {{x_p}{\rm{,\;}}{{\dot x}_p}{\rm{,\;}}\kappa \left( \theta  \right)} \right)$ that use states derivatives in addition to states themselves. Despite that fact, in this study we focus our attention on control laws with additional feedbacks from the state ${x_p}\left( t \right)$ only.

The equation of the control law ${u^*}\left( t \right)$ and the one to calculate its parameters $\kappa \left( \theta  \right)$ from \cite{b6} motivate to introduce the following assumption.

\textbf{Assumption 3.} \emph{There exist continuous mappings ${\cal G}\left( \theta  \right){\rm{:\;}}{\mathbb{R}^{{n_\theta }}} \to {\mathbb{R}^{{n_\kappa } \times {n_\kappa }}}$, ${\cal S}\left( \theta  \right){\rm{:\;}}{\mathbb{R}^{{n_\theta }}} \to {\mathbb{R}^{{n_\kappa}}}$ such that for all ${{\cal M}_\theta }\left( t \right) \ge 0$ the following equations hold:}
\begin{equation}\label{eq11}
\begin{gathered}
  \mathcal{S}\left( \theta  \right) = \mathcal{G}\left( \theta  \right)\kappa \left( \theta  \right){\text{,}} \\ 
  {\Pi _\kappa }\left( {{\mathcal{M}_\theta }} \right)\mathcal{G}\left( \theta  \right) \!=\! {\mathcal{T}_\mathcal{G}}\left( {{\Xi _\mathcal{G}}\left( {{\mathcal{M}_\theta }} \right)\theta } \right){\text{:\;}}{\mathbb{R}^{{\mathcal{M}_\theta }_\mathcal{G}}} \to {\mathbb{R}^{{n_\kappa } \times {n_\kappa }}}{\text{,}} \\ 
  {\Pi _\kappa }\left( {{\mathcal{M}_\theta }} \right)\mathcal{S}\left( \theta  \right) = {\mathcal{T}_\mathcal{S}}\left( {{\Xi _\mathcal{S}}\left( {{\mathcal{M}_\theta }} \right)\theta } \right){\text{:\;}}{\mathbb{R}^{{\mathcal{M}_\theta }_\mathcal{S}}} \to {\mathbb{R}^{{n_\kappa }}}{\text{,}} \\ 
\end{gathered}
\end{equation}
\emph{where}
\begin{equation}\label{eq12}
\begin{gathered}
{\rm{det}}\left\{ {{\Pi _\kappa }\left( {{{\cal M}_\theta }} \right)} \right\} \!\ge\! {\cal M}_\theta ^{{\ell _\kappa }}\left( t \right){\rm{,\;}}{\ell _\kappa } \!\ge\! 1{\rm{, rank}}\left\{ {{\cal G}\left( \theta  \right)} \right\} \!=\! {n_\kappa }{\rm{,\;}}\\
{\Xi _{\left( . \right)}}\left( {{{\cal M}_\theta }} \right) = {{\overline \Xi }_{\left( . \right)}}\left( {{{\cal M}_\theta }} \right){{\cal M}_\theta }\left( t \right) \in {\mathbb{R}^{{{\cal M}_\theta }_{\left( . \right)} \times {n_\theta }}}{\rm{,\;}}\\
{\Xi _{\left( . \right)}}_{ij}\left( {{{\cal M}_\theta }} \right) = c{\cal M}_\theta ^\ell \left( t \right){\rm{,\;}}c \in \left\{ {0,{\rm{ 1}}} \right\}{\rm{,\;}}\ell  > 0{\rm{,}}
\end{gathered}
\end{equation}
\emph{and all introduced mappings are known}.

Assumption 2 describes necessary and sufficient conditions to transform the regression equation with respect to (w.r.t.) $\theta $ into the one w.r.t. the ideal control law parameters. The parameters of all ideal control laws from \cite{b6} satisfy this assumption. For example, the equations to obtain parameters of the baseline PI-controller \cite[Subsection B, Equation (14)]{b6}, is considered:
\begin{equation}\label{eq13}
\begin{gathered}
{K_{\rm{P}}}\left( \theta  \right) = 2\sqrt {\frac{{{\theta _1}}}{{{\theta _3}}}} {\rm{,\;}}{K_{\rm{I}}}\left( \theta  \right) = \frac{{{\theta _1}}}{{{\theta _2}{\theta _3}}}{\rm{.}}
\end{gathered}
\end{equation}

Then the mappings ${\cal S}\left( \theta  \right){\rm{,\;}}{\cal G}\left( \theta  \right)$ are written as:
\begin{displaymath}
\begin{gathered}
{\cal G}\left( \theta  \right) = {\begin{bmatrix}
{\sqrt {{\theta _3}} }&0\\
0&{{\theta _2}{\theta _3}}
\end{bmatrix}}{\rm{,\;}}{\cal S}\left( \theta  \right) = {\begin{bmatrix}
{2\sqrt {{\theta _1}} }\\
{{\theta _1}}
\end{bmatrix}}
\end{gathered}
\end{displaymath}
and the transformations ${\Pi _\kappa }\left( {{{\cal M}_\theta }} \right){\rm{,\;}}{{\cal T}_{\cal S}}\left( . \right){\rm{,\;}}{{\cal T}_{\cal G}}\left( . \right)$ take the form:
\begin{displaymath}
\begin{gathered}
{\Pi _\kappa }\left( {{{\cal M}_\theta }} \right) = {\begin{bmatrix}
{{\cal M}_\theta ^{0.5}}&0\\
0&{{\cal M}_\theta ^2}
\end{bmatrix}}{\rm{,\;}}\\
{{\cal T}_{\cal G}}\left( {{{\overline \Xi }_{\cal G}}\left( {{{\cal M}_\theta }} \right){{\cal Y}_\theta }} \right) = {\begin{bmatrix}
{\sqrt {{{\cal Y}_{\theta 3}}} }&0\\
0&{{{\cal Y}_{\theta 2}}{{\cal Y}_{\theta 3}}}
\end{bmatrix}}{\rm{,}}\\
{{\cal T}_{\cal S}}\left( {{{\overline \Xi }_{\cal S}}\left( {{{\cal M}_\theta }} \right){{\cal Y}_\theta }} \right) = {\begin{bmatrix}
{2\sqrt {{{\cal Y}_{\theta 1}}} }&0\\
0&{{{\cal M}_\theta }{{\cal Y}_{\theta 1}}}
\end{bmatrix}},
\end{gathered}
\end{displaymath}
which allow one to transform the regression equation \linebreak ${{\cal Y}_\theta }\left( t \right) = \Delta \left( t \right)\theta $ w.r.t. $\theta $ into the one w.r.t. $\kappa \left( \theta  \right)$ with measurable regressor and regressand ({\it i.e.} measurable regression equation):
\begin{equation}\label{eq14}
\begin{gathered}
{{\cal T}_{\cal S}}\left( {{{\overline \Xi }_{\cal S}}\left( {{{\cal M}_\theta }} \right){{\cal Y}_\theta }} \right) = {{\cal T}_{\cal G}}\left( {{{\overline \Xi }_{\cal G}}\left( {{{\cal M}_\theta }} \right){{\cal Y}_\theta }} \right)\kappa \left( \theta  \right).
\end{gathered}
\end{equation}

Here a square root operation is a safe one as it is {\it a priori} known that $\theta  \in \mathbb{R}_{^ > }^3$. 

In general case, the equations to calculate the control law \eqref{eq10} parameters can be obtained with the help of the pole placement design:
\begin{equation}\label{eq15}
\begin{gathered}
{\rm{det}}\left\{ {s{I_4} - \left( {{\cal A}\left( \theta  \right) + {\cal B}\left( \theta  \right){\kappa ^{\rm{T}}}\left( \theta  \right)} \right)} \right\}= \;\;\;\;\;\;\;\;\;\;\;\;\;\;\;\;\;\;\;\;\;\\\hfill
\;\;\;\;\;\;\;\;\;\;\;\;\;\;\;\;\;\;\;\;\;\;\;\;\;\;\;\;\;\;={\left( {{s^2} + 2{\xi _d}{\omega _d}s + \omega _d^2} \right)^2}{\rm{,}}
\end{gathered}
\end{equation}
where ${\omega _d}$ and ${\xi _d}$ denote required resonant frequency and damping coefficient, respectively. 

Having solved \eqref{eq15} w.r.t. the elements of the vector $\kappa \left( \theta  \right)$, we obtain the following equations to calculate the parameters of \eqref{eq10}:
\begin{equation}\label{eq16}
\begin{gathered}
\kappa \left( \theta  \right) = { {\begin{bmatrix}
{{K_I}\left( \theta  \right)}&{{K_P}\left( \theta  \right)}&{{K_{1x}}\left( \theta  \right)}&{{K_{2x}}\left( \theta  \right)}
\end{bmatrix}}^{\rm{T}}}{\rm{,}}\\
{K_I}\left( \theta  \right) = {\theta _1}{\theta _2}{\theta _3}\omega _d^4{\rm{,}}\\
{K_P}\left( \theta  \right) =  - 4{\theta _1}{\xi _d}{\omega _d}{\rm{,}}\\
{K_{1x}}\left( \theta  \right) =  - 4{\theta _1}{\xi _d}{\omega _d}\left( {{\theta _2}{\theta _3}\omega _d^2 - 1} \right){\rm{,}}\\
{K_{2x}}\left( \theta  \right) = {\textstyle{{ - {\theta _1}{\theta _2}{\theta _3}\omega _d^2\left( {4\xi _d^2 - {\theta _2}{\theta _3}\omega _d^2 + 2} \right) + {\theta _2} + {\theta _1}} \over {{\theta _2}}}}.
\end{gathered}
\end{equation}

Then the mappings ${\cal G}\left( \theta  \right),\;{\cal S}\left( \theta  \right)$ take the form:
\begin{displaymath}
\begin{gathered}
{\cal G}\left( \theta  \right) = diag\left\{ {1{\rm{,\;1,\;1,\;}}{\theta _2}} \right\}{\rm{,\;}}\\
{\cal S}\left( \theta  \right) = col\left\{ \!\begin{array}{c}
{\theta _1}{\theta _2}{\theta _3}\omega _d^4{\rm{,\;}}\\
 - 4{\theta _1}{\xi _d}{\omega _d}{\rm{,\;}}\\
 - 4{\theta _1}{\xi _d}{\omega _d}\left( {{\theta _2}{\theta _3}\omega _d^2 - 1} \right){\rm{,\;}}\\
 - {\theta _1}{\theta _2}{\theta _3}\omega _d^2\left( {4\xi _d^2 - {\theta _2}{\theta _3}\omega _d^2 + 2} \right) + {\theta _2} + {\theta _1}
\end{array} \!\right\}{\rm{,}}
\end{gathered}
\end{displaymath}
and ${\Pi _\kappa }\left( {{{\cal M}_\theta }} \right){\rm{,\;}}{{\cal T}_{\cal S}}\left( . \right){\rm{,\;}}{{\cal T}_{\cal G}}\left( . \right)$ are written as:
\begin{displaymath}
\begin{gathered}
{\Pi _\kappa }\left( {{{\cal M}_\theta }} \right) = diag\left\{ {{\cal M}_\theta ^3{\rm{,\;}}{{\cal M}_\theta }{\rm{,\;}}{\cal M}_\theta ^3{\rm{,\;}}{\cal M}_\theta ^5} \right\}{\rm{,}}\\
{\rm{ }}{{\cal T}_{\cal G}}\left( {{{\overline \Xi }_{\cal G}}\left( {{{\cal M}_\theta }} \right){{\cal Y}_\theta }} \right) = diag\left\{ {{\cal M}_\theta ^3{\rm{,\;}}{{\cal M}_\theta }{\rm{,\;}}{\cal M}_\theta ^3{\rm{,\;}}{\cal M}_\theta ^4{{\cal Y}_{\theta 2}}} \right\}{\rm{,}}\\
{{\cal T}_{\cal S}}\left( {{{\overline \Xi }_{\cal S}}\left( {{{\cal M}_\theta }} \right){{\cal Y}_\theta }} \right) \!\!=\!\! col\left\{ \!\!\!\begin{array}{c}
{{\cal Y}_{\theta 1}}{{\cal Y}_{\theta 2}}{{\cal Y}_{\theta 3}}\omega _d^4{\rm{,\;}}\\
 - 4{{\cal Y}_{\theta 1}}{\xi _d}{\omega _d}{\rm{,\;}}\\
 - 4{{\cal Y}_{\theta 1}}{\xi _d}{\omega _d}\left( {{{\cal Y}_{\theta 2}}{{\cal Y}_{\theta 3}}\omega _d^2 - {\cal M}_\theta ^2} \right){\rm{,\;}}\\
 - {{\cal Y}_{\theta 1}}{{\cal Y}_{\theta 2}}{{\cal Y}_{\theta 3}}\omega _d^2( {4{\cal M}_\theta ^2\xi _d^2 -}\\{- {{\cal Y}_{\theta 2}}{{\cal Y}_{\theta 3}}\omega _d^2 + 2{\cal M}_\theta ^2}) +\\+ {\cal M}_\theta ^4{{\cal Y}_{\theta 2}} + {\cal M}_\theta ^4{{\cal Y}_{\theta 1}}
\end{array}\!\!\! \right\},
\end{gathered}
\end{displaymath}
which allows one to transform the regression equation ${{\cal Y}_\theta }\left( t \right) = {{\cal M}_\theta }\left( t \right)\theta $ w.r.t. $\theta $ into the measurable one \eqref{eq14}.

By analogy with the above-presented derivations, an interested reader can rewrite other control laws from \cite{b6} as \eqref{eq14}. Therefore, Assumption 2 is not restrictive for the considered system \eqref{eq4} and control law \eqref{eq10}. 

As the system parameters $\theta $ are unknown and the state vector ${x_p}\left( t \right)$ is not measured, the following adaptive control problem is stated.

\textbf{Goal.} Let Assumptions 1-3 be met, then the aim is to obtain an adaptive control law in the following form:
\begin{equation}\label{eq18}
\begin{gathered}
u\left( t \right) = {\hat \kappa ^{\rm{T}}}\left( t \right)\hat x\left( t \right) = {\hat K_{{\rm{PI}}}}\left( t \right){e_y}\left( t \right) + {\hat K_{x} }\left( t \right){\hat x_p}\left( t \right){\rm{,}}
\end{gathered}
\end{equation}
which provides for $\zeta \left( t \right) = {{\begin{bmatrix}
{\tilde x_p^{\rm{T}}\left( t \right)}&{e_{ref}^{\rm{T}}\left( t \right)}&{{{\tilde \kappa }^{\rm{T}}}\left( t \right)}
\end{bmatrix}}^{\rm{T}}}$ that:
\begin{equation}\label{eq19}
\begin{gathered}
\mathop {{\rm{lim}}}\limits_{t \to \infty } \left\| {\zeta \left( t \right)} \right\| = 0{\rm{ }}\left( {{\rm{exp}}} \right){\rm{,}}
\end{gathered}
\end{equation}
where $\hat \kappa \left( t \right)$ is an estimate of the full-information-based control law \eqref{eq10} parameters, ${\tilde x_p}\left( t \right) = {\hat x_p}\left( t \right) - {x_p}\left( t \right)$ stands for the plant states observation error, ${e_{ref}}\left( t \right) = x\left( t \right) - {x^*}\left( t \right)$ denotes tracking error of states ${x^*}\left( t \right)$ of a system with ideal control law \eqref{eq10}.
\section{Main result}
The states ${x_p}\left( t \right)$ and control law parameters $\kappa \left( \theta  \right)$ depend on unknown physical parameters $\theta $ of the system. The first aim is to obtain a regression equation that relates the parameters $\theta $ to the functions of measurable signals $y\left( t \right){\rm{,\;}}u\left( t \right)$:
\begin{equation}\label{eq20}
\begin{gathered}
{{\cal Y}_\theta }\left( t \right) = {{\cal M}_\theta }\left( t \right)\theta ,\\
{{\cal Y}_\theta }\left( t \right) = {f_{\cal Y}}\left( {y{\rm{,\;}}u} \right){\rm{,\;}}{{\cal M}_\theta }\left( t \right) = {f_{\cal M}}\left( {y{\rm{,\;}}u} \right).
\end{gathered}
\end{equation}

In order to achieve it, the model \eqref{eq4} is represented in the observer canonical form. To this end, a linear transformation $\xi \left( t \right) = T\left( \theta  \right){x_p}\left( t \right)$ is introduced, and, according to \cite[p. 263]{b30}, a nonsingular matrix $T\left( \theta  \right)$ is defined as follows:
\begin{displaymath}
    {T_I}\left( \theta  \right){\rm{:}} = {T^{ - 1}}\left( \theta  \right) = {\begin{bmatrix}
{{A^2}\left( \theta  \right){{\cal O}_3}\left( \theta  \right)}&{A\left( \theta  \right){{\cal O}_3}\left( \theta  \right)}&{{{\cal O}_3}\left( \theta  \right)}
\end{bmatrix}} {\rm{,}}
\end{displaymath}
\vspace{-0.5cm}
\begin{equation}\label{eq21}
\begin{gathered}
{{\cal O}_3}\left( \theta  \right) = {\cal O}\left( \theta  \right){{\begin{bmatrix}
0&0&1
\end{bmatrix}} ^{\rm{T}}}{\rm{,}}\\
{{\cal O}^{ - 1}}\left( \theta  \right) = {{\begin{bmatrix}
C&{{{\left( {A\left( \theta  \right)} \right)}^{\rm{T}}}C}&{{{\left( {{A^2}\left( \theta  \right)} \right)}^{\rm{T}}}C}
\end{bmatrix}}^{\rm{T}}}{\rm{,}}
\end{gathered}
\end{equation}
then a differential equation for $\xi \left( t \right)$ is written as:
\begin{equation}\label{eq22}
\begin{gathered}
\dot \xi \left( t \right) = {A_0}\xi \left( t \right) + \hfill \\ \hfill  +{\psi _a}\left( \theta  \right)y\left( t \right) + {\psi _b}\left( \theta  \right)u\left( t \right)
+ {\psi _d}\left( \theta  \right)\delta \left( t \right) = \\ = {A_0}\xi \left( t \right) + {\phi ^{\rm{T}}}\left( {y{\rm{,\;}}u{\rm{,\;}}\delta } \right)\eta \left( \theta  \right){\rm{,}}\hfill\\
y\left( t \right) = C_0^{\rm{T}}\xi \left( t \right){\rm{,\;}}\xi \left( {{t_0}} \right) = {\xi _0}\left( \theta  \right) = T\left( \theta  \right){x_0}{\rm{,}}\hfill
\end{gathered}
\end{equation}
where
\begin{equation}\label{eq23}
\begin{gathered}
  {A_0} = {\begin{bmatrix}
  0&1&0 \\ 
  0&0&1 \\ 
  0&0&0 
\end{bmatrix}},\quad C_0^{\text{T}} = {\begin{bmatrix}
  1&0&0 
\end{bmatrix}}, \\ 
  {\psi _a}\left( \theta  \right) = T\left( \theta  \right)A\left( \theta  \right){T_I}\left( \theta  \right){C_0} = {{\begin{bmatrix}
  0&{ - \tfrac{{{\theta _1} + {\theta _2}}}{{{\theta _1}{\theta _2}{\theta _3}}}}&0 
\end{bmatrix}}^{\text{T}}}{\text{,}} \\ 
  {\text{ }}{\psi _b}\left( \theta  \right) = T\left( \theta  \right)B\left( \theta  \right) = { {\begin{bmatrix}
  {\theta _1^{ - 1}}&0&{\tfrac{1}{{{\theta _1}{\theta _2}{\theta _3}}}} 
\end{bmatrix}}^{\text{T}}}{\text{, }}\\
{\psi _d}\left( \theta  \right) = T\left( \theta  \right)D\left( \theta  \right) = {{\begin{bmatrix}
  0&0&{ - \tfrac{1}{{{\theta _1}{\theta _2}{\theta _3}}}} 
\end{bmatrix}} ^{\text{T}}}{\text{,}} \\ 
  \eta \left( \theta  \right) = {{\begin{bmatrix}
  {\psi _a^{\text{T}}\left( \theta  \right)}&{\psi _b^{\text{T}}\left( \theta  \right)}&{\psi _d^{\text{T}}\left( \theta  \right)} 
\end{bmatrix}}^{\text{T}}}{\text{,}} \\ 
  {T_I}\left( \theta  \right) = {\begin{bmatrix}
  1&0&0 \\ 
  { - {\theta _1}\theta _2^{ - 1}}&0&{{\theta _1}{\theta _3}} \\ 
  0&{ - {\theta _1}}&0 
\end{bmatrix}} {\text{, }}\\
{\phi ^{\text{T}}}\left( . \right) = {\begin{bmatrix}
  0&0&0&u&0&0&0&0&0 \\ 
  0&y&0&0&0&0&0&0&0 \\ 
  0&0&0&0&0&u&0&0&\delta  
\end{bmatrix}}. \\ 
\end{gathered}
\end{equation}

So, using \eqref{eq22}, \eqref{eq23} and the results of \cite{b25,b31}, the following parametrization is obtained.

\textbf{Lemma 1.} \emph{Let ${t_\epsilon} > {t_0}$ be a sufficiently large predefined time instance, then for all $t \geq {t_\epsilon}$ the unknown parameters $\eta \left( \theta  \right)$ satisfy the regression equation:}
\begin{equation}\label{eq24}
\begin{gathered}
{\cal Y}\left( t \right) = \Delta \left( t \right)\eta \left( \theta  \right){\rm{,}}\\
{\cal Y}\left( t \right) = 
\begin{bmatrix}
{k\left( t \right) \cdot {\rm{adj}}\left\{ {\varphi \left( t \right)} \right\}q\left( t \right)}\\
{{{\cal Y}_{{\psi _d}}}\left( t \right)}
\end{bmatrix}{\rm{,}}
\\
{\Delta \left( t \right) = k\left( t \right) \cdot {\rm{det}}\left\{ {\varphi \left( t \right)} \right\}{\rm{,}}}\\
{{{\cal L}_d} = {\begin{bmatrix}
{{0_{3 \times 5}}}&{\begin{matrix}
{{0_2}}\\
{ - 1}
\end{matrix}}
\end{bmatrix}}},
\\
{{\cal Y}_{{\psi _d}}}\left( t \right) = k\left( t \right) \cdot {{\cal L}_d}{\rm{adj}}\left\{ {\varphi \left( t \right)} \right\}q\left( t \right) = \Delta \left( t \right){\psi _d}\left( \theta  \right){\rm{,}}
\end{gathered}
\end{equation}
\emph{where}
\begin{equation}\label{eq25}
\begin{gathered}
  q\left( t \right) = \int\limits_{{t_\epsilon}}^t {e^{ - \sigma \left( {\tau  - {t_\epsilon}} \right)}}{{\overline \varphi }_f}\left( \tau  \right)( \overline q\left( \tau  \right) - {k_1}{{\overline q}_f}\left( \tau  \right) -\\- {\beta ^{\text{T}}}\left( {{F_f}\left( \tau  \right) + l{y_f}\left( \tau  \right)} \right) )d\tau  {\text{, }}q\left( {{t_\epsilon}} \right) = {0_{{\text{2}}n}}, \\ 
  \varphi \left( t \right) = \int\limits_{{t_\epsilon}}^t {{e^{ - \sigma \left( {\tau  - {t_\epsilon}} \right)}}{{\overline \varphi }_f}\left( \tau  \right)\overline \varphi _f^{\text{T}}\left( \tau  \right)d\tau } {\text{, }}\\
  \varphi \left( {{t_\epsilon}} \right) = {0_{{\text{2}}n \times {\text{2}}n}}, \\ 
\end{gathered}
\end{equation}

\begin{equation}\label{eq26}
\begin{gathered}
{{\dot {\overline q}}_f}\left( t \right) =  - {k_1}{{\overline q}_f}\left( t \right) + \overline q\left( t \right){\rm{,\;}}{{\overline q}_f}\left( {{t_0}} \right) = 0,\\
{{\dot {\overline \varphi} }_f}\left( t \right) =  - {k_1}{{\overline \varphi }_f}\left( t \right) + \overline \varphi \left( t \right){\rm{,\;}}{{\overline \varphi }_f}\left( {{t_0}} \right) = {0_{{\rm{2}}n}},\\
{{\dot F}_f}\left( t \right) =  - {k_1}{F_f}\left( t \right) + F\left( t \right){\rm{,\;}}{F_f}\left( {{t_0}} \right) = {0_{{n_\delta }}},\\
{{\dot y}_f}\left( t \right) =  - {k_1}{y_f}\left( t \right) + y\left( t \right){\rm{,\;}}{y_f}\left( {{t_0}} \right) = 0,
\end{gathered}
\end{equation}

\begin{equation}\label{eq27}
\begin{gathered}
  \overline q\left( t \right) = y\left( t \right) - C_0^{\text{T}}z{\text{, }}\overline \varphi \left( t \right) = {\begin{bmatrix}
  {{{\dot \Omega }^{\text{T}}}{C_0} + {N^{\text{T}}}\beta } \\ 
  {{{\dot P}^{\text{T}}}{C_0} + {H^{\text{T}}}\beta } 
\end{bmatrix}}{\text{,}} \\ 
  \dot z\left( t \right) = {A_K}z\left( t \right) + Ky\left( t \right){\text{, }}z\left( {{t_0}} \right) = {0_n}{\text{,}} \\ 
  \dot \Omega \left( t \right) = {A_K}\Omega \left( t \right) + {I_n}y\left( t \right){\text{, }}\Omega \left( {{t_0}} \right) = {0_{n \times n}}{\text{,}} \\ 
  \dot P\left( t \right) = {A_K}P\left( t \right) + {I_n}u\left( t \right){\text{, }}P\left( {{t_0}} \right) = {0_{n \times n}}{\text{,}} \\ 
  \dot F\left( t \right) = GF\left( t \right) + Gly\left( t \right) - lC_0^{\text{T}}\dot z\left( t \right){\text{, }}F\left( {{t_0}} \right) = {0_{{n_\delta }}}{\text{,}} \\ 
  \dot H\left( t \right) = GH\left( t \right) - lC_0^{\text{T}}\dot P\left( t \right){\text{, }}H\left( {{t_0}} \right) = {0_{{n_\delta } \times n}}{\text{,}} \\ 
  \dot N\left( t \right) = GN\left( t \right) - lC_0^{\text{T}}\dot \Omega \left( t \right){\text{, }}N\left( {{t_0}} \right) = {0_{{n_\delta } \times n}}{\text{,}} \\ 
\end{gathered}
\end{equation}
\emph{and, if $\overline \varphi \left( t \right) \in {\rm{FE}}$ over the time range $\left[ {{t_\epsilon}{\text{;\;}}{t_e}} \right]$, then for all $t \ge {t_e}$ it holds that $\Delta \left( t \right) \ge {\Delta _{{\rm{min}}}} > 0$.}

\emph{Here $k\left( t \right) > 0$ is a time-varying (or time-invariant) amplifier, ${k_1} > 0{\rm{,\;}}\sigma  > 0$ stands for filters time constants, ${A_K} = {A_0} - KC_0^{\rm{T}}{\rm{,\;}}G$ denote stable matrices of appropriate dimension, $l \in {\mathbb{R}^{{n_\delta }}}$ is a vector such that the pair $\left( {G{\rm{,\;}}l} \right)$ is controllable and $G$ is chosen to meet $\sigma \left\{ {{{\cal A}_\delta }} \right\} \cap \sigma \left\{ G \right\} = 0$, $\beta  \in {\mathbb{R}^{{n_\delta }}}$ is a solution of a set of equations:}
\begin{displaymath}
\begin{gathered}
{M_\delta }{{\cal A}_\delta } - G{M_\delta } = l\overline h_\delta ^{\rm{T}}{\rm{,\;}}\overline h_\delta ^{\rm{T}} = h_\delta ^{\rm{T}}{{\cal A}_\delta }{\rm{,}}\\
\beta  = \overline h_\delta ^{\rm{T}}M_\delta ^{ - 1}. 
\end{gathered}
\end{displaymath}

\emph{Proof of Lemma 1 is postponed to Appendix}.

~

It should be noted that in \eqref{eq24} the functions of measurable signals ${\cal Y}\left( t \right)$ and $\Delta \left( t \right)$ relate to the nonlinear functions $\eta \left( \theta  \right)$ of the unknown parameters $\theta $. In order to obtain linear relation \eqref{eq20} between $\theta $ and some functions ${f_{\cal Y}}\left( {y{\rm{,\;}}u} \right){\rm{,\;}}{f_{\cal M}}\left( {y{\rm{,\;}}u} \right)$ of measurable signals, $\theta $ are to be expressed from \eqref{eq24}.

To this end, the following function is introduced:
\begin{equation}\label{eq28}
\begin{gathered}
{\psi _{ab}}\left( \theta  \right) = {{\cal L}_{ab}}\eta \left( \theta  \right) =\\= \!\!{\begin{bmatrix}
{\begin{array}{*{20}{c}}
0&0&0&1&0&0\\
0&1&0&0&0&0\\
0&0&0&0&0&1
\end{array}}&{{0_{3 \times 3}}}
\end{bmatrix}}\eta \left( \theta  \right) = {\begin{bmatrix}
{\theta _1^{ - 1}}\\
{ - {\textstyle{{{\theta _1} + {\theta _2}} \over {{\theta _1}{\theta _2}{\theta _3}}}}}\\
{{\textstyle{1 \over {{\theta _1}{\theta _2}{\theta _3}}}}}
\end{bmatrix}},
\end{gathered}
\end{equation}
such that for all $\theta  \in \mathbb{R}_{^ > }^3$ we meet the existence condition of a function $\theta  = {\cal F}\left( {{\psi _{ab}}} \right)$, which is an inverse one to ${\psi _{ab}}\left( \theta  \right)$: 
\begin{equation}\label{eq29}
\begin{gathered}
{\rm{de}}{{\rm{t}}^2}\left\{ {{\nabla _\theta }{\psi _{ab}}\left( \theta  \right)} \right\} =
\end{gathered}
\end{equation}
\vspace{-15pt}
\begin{gather*}
= {\rm{de}}{{\rm{t}}^2}\!\left\{\! {{\begin{bmatrix}
{ - \theta _1^{ - 2}}&0&0\\
{\theta _1^{ - 2}\theta _3^{ - 1}}&{\theta _2^{ - 2}\theta _3^{ - 1}}&{{\textstyle{{{\theta _1} + {\theta _2}} \over {{\theta _1}{\theta _2}\theta _3^2}}}}\\
{ - \theta _1^{ - 2}\theta _2^{ - 1}\theta _3^{ - 1}}&{ - \theta _1^{ - 1}\theta _2^{ - 2}\theta _3^{ - 1}}&{ - \theta _1^{ - 1}\theta _2^{ - 1}\theta _3^{ - 2}}
\end{bmatrix}}} \!\right\} \!=\\
={\left( { - \theta _1^{ - 4}\theta _2^{ - 2}\theta _3^{ - 3}} \right)^2} > 0. 
\end{gather*}
Owing to \eqref{eq29}, the parameters $\theta $ are expressed from \eqref{eq28}:
\begin{equation}\label{eq30}
\begin{gathered}
{\theta _1} = \psi _{1ab}^{ - 1},\\
{\theta _2} = {\textstyle{{ - \psi _{1ab}^{ - 1}{\psi _{3ab}} - {\psi _{2ab}}} \over {{\psi _{3ab}}}}},\\
{\theta _3} = {\textstyle{1 \over {\psi _{1ab}^{ - 1}\left( { - \psi _{1ab}^{ - 1}{\psi _{3ab}} - {\psi _{2ab}}} \right)}}}.
\end{gathered}
\end{equation}

Equation \eqref{eq30} is rewritten as:
\begin{equation}\label{eq31}
\begin{gathered}
{\cal W}\left( {{\psi _{ab}}} \right) = {\cal R}\left( {{\psi _{ab}}} \right)\theta {\rm{,}}\\
{\cal W}\left( {{\psi _{ab}}} \right) = col\left\{ {1{\rm{,\;}} - {\psi _{3ab}} - {\psi _{1ab}}{\psi _{2ab}}{\rm{,\;}}\psi _{1ab}^2} \right\}{\rm{,}}\\
{\cal R}\left( {{\psi _{ab}}} \right) \!=\! diag\!\left\{ {{\psi _{1ab}}{\rm{,\;}}{\psi _{3ab}}{\psi _{1ab}}{\rm{,\;}} - {\psi _{3ab}} - {\psi _{1ab}}{\psi _{2ab}}} \right\},
\end{gathered}
\end{equation}
then, having multiplied \eqref{eq31} by 
\begin{gather*}
{\Pi _\theta }\left( {\Delta \left( t \right)} \right) = {\rm{diag}}\left\{ {\Delta \left( t \right){\rm{,\;}}{\Delta ^2}\left( t \right){\rm{,\;}}{\Delta ^2}\left( t \right)} \right\}    
\end{gather*} 
and substituted \eqref{eq24}, it is obtained that $\left( {{{\cal Y}_{ab}}\left( t \right) = {{\cal L}_{ab}}{\cal Y}\left( t \right)} \right)$:
\begin{equation}\label{eq32}
\begin{gathered}
{{\cal T}_{\cal W}}\left( {{{\overline \Xi }_{\cal W}}\left( \Delta  \right){{\cal Y}_{ab}}} \right) = {{\cal T}_{\cal R}}\left( {{{\overline \Xi }_{\cal R}}\left( {\Delta \left( t \right)} \right){{\cal Y}_{ab}}} \right)\theta {\rm{,}}\\
\end{gathered}
\end{equation}
\vspace{-20pt}
\begin{gather*}
{{\cal T}_{\cal W}}\left( {{{\overline \Xi }_{\cal W}}\left( \Delta  \right){{\cal Y}_{ab}}} \right) = col\left\{ {\Delta {\rm{,\;}} - \Delta {{\cal Y}_{3ab}} - {{\cal Y}_{1ab}}{{\cal Y}_{2ab}}{\rm{,\;}}{\cal Y}_{1ab}^2} \right\}{\rm{,}}\\
{{\cal T}_{\cal R}}\left( {{{\overline \Xi }_{\cal R}}\left( {\Delta \left( t \right)} \right){{\cal Y}_{ab}}} \right) =\\
= diag\left\{ {{{\cal Y}_{1ab}}{\rm{,\;}}{{\cal Y}_{3ab}}{{\cal Y}_{1ab}}{\rm{,\;}} - \Delta {{\cal Y}_{3ab}} - {{\cal Y}_{1ab}}{{\cal Y}_{2ab}}} \right\}{\rm{,}}    
\end{gather*}
and, having multiplied \eqref{eq32} by 
\begin{gather*}
   {\rm{det}}\left\{ {{{\cal T}_{\cal R}}\left( {{{\overline \Xi }_{\cal R}}\left( {\Delta \left( t \right)} \right){{\cal Y}_{ab}}} \right)} \right\}{\rm{adj}}\left\{ {{{\cal T}_{\cal R}}\left( {{{\overline \Xi }_{\cal R}}\left( {\Delta \left( t \right)} \right){{\cal Y}_{ab}}} \right)} \right\}, 
\end{gather*}
 we have the required regression equation of the form (20):
\begin{equation}\label{eq33}
\begin{gathered}
{{\cal Y}_\theta }\left( t \right) = {{\cal M}_\theta }\left( t \right)\theta, \\
{{\cal Y}_\theta }\left( t \right) = {\rm{det}}\left\{ {{{\cal T}_{\cal R}}\left( {{{\overline \Xi }_{\cal R}}\left( {\Delta \left( t \right)} \right){{\cal Y}_{ab}}} \right)} \right\}\times\hfill\\\hfill
\times{\rm{adj}}\left\{ {{{\cal T}_{\cal R}}\left( {{{\overline \Xi }_{\cal R}}\left( {\Delta \left( t \right)} \right){{\cal Y}_{ab}}} \right)} \right\}{{\cal T}_{\cal W}}\left( {{{\overline \Xi }_{\cal W}}\left( \Delta  \right){{\cal Y}_{ab}}} \right){\rm{,}}\\
{{\cal M}_\theta }\left( t \right) = {\rm{de}}{{\rm{t}}^2}\left\{ {{{\cal T}_{\cal R}}\left( {{{\overline \Xi }_{\cal R}}\left( {\Delta \left( t \right)} \right){{\cal Y}_{ab}}} \right)} \right\}{\rm{,}}
\end{gathered}
\end{equation}
where, as
\begin{displaymath}
\begin{gathered}
{{\cal M}_\theta }\left( t \right) = {\rm{de}}{{\rm{t}}^2}\left\{ {{{\cal T}_{\cal R}}\left( {{{\overline \Xi }_{\cal R}}\left( {\Delta \left( t \right)} \right){{\cal Y}_{ab}}} \right)} \right\} =\;\;\;\;\;\;\;\;\;\;\;\;\;\;\;\;\;\;\\
\;\;\;\;\;\;\;\;\;\;\;\;\;\;\;\;\;\;\;\;\;= \underbrace {{\rm{de}}{{\rm{t}}^2}\left\{ {{\cal R}\left( {{\psi _{ab}}} \right)} \right\}}_{ > 0}\underbrace {{\rm{de}}{{\rm{t}}^2}\left\{ {{\Pi _\theta }\left( {\Delta \left( t \right)} \right)} \right\}}_{ = {\Delta ^{10}}\left( t \right)}{\rm{,}}
\end{gathered}
\end{displaymath}
if $\overline \varphi \left( t \right) \in {\rm{FE}}$ over $\left[ {{t_\epsilon}{\text{; }}{t_e}} \right]$ , then for all $t \ge {t_e}$ it holds that $\left| {{{\cal M}_\theta }\left( t \right)} \right| \ge \underline {{{\cal M}_\theta }}  > 0$.

Based on the results of Lemma 1 and equation \eqref{eq33}, the solution of the stated problem can be divided into two main steps. At the first one, in order to implement additional feedback ${K_{x}}\left( \theta  \right){x_p}\left( t \right)$, it is necessary to obtain estimates of states ${x_p}\left( t \right)$ with the help of the measurable regression equation \eqref{eq33}. At the second step, it is necessary to transform the regression equation w.r.t. $\theta $ into the one w.r.t. the parameters $\kappa \left( \theta  \right)$ and derive the law of their identification.

Thus, both the states ${x_p}\left( t \right)$ observer and the law of controller parameters $\kappa \left( \theta  \right)$ identification will be derived on the basis of the measurable regression equation \eqref{eq33} w.r.t. parameters $\theta $, which is obtained by transformations \eqref{eq24}-\eqref{eq27}, \eqref{eq32}, \eqref{eq33} and has a bounded from zero scalar regressor ${{\cal M}_\theta }\left( t \right)$ for all $t \ge {t_e}$.

\subsection{States Observation}

The solution of the exosystem \eqref{eq6} equation is written as:
\begin{equation}\label{eq34}
\begin{gathered}
{{\dot \Phi }_\delta }\left( t \right) = {{\cal A}_\delta }{\Phi _\delta }\left( t \right){\rm{,\;}}{\Phi _\delta }\left( {{t_0}} \right) = {I_{{n_\delta }}}{\rm{,}}\hfill\\
{x_\delta }\left( t \right) = {\Phi _\delta }\left( t \right){x_{\delta 0}}{\rm{,}}\hfill
\end{gathered}
\end{equation}
and hence estimates of the unmeasurable states of system \eqref{eq4} can be formed according to the following set of equations:
\begin{equation}\label{eq35}
\begin{gathered}
{{\hat x}_p}\left( t \right) = {{\hat T}_I}\left( t \right)\hat \xi \left( t \right){\rm{,}}\hfill\\
\dot {\hat \xi} \left( t \right) = {A_0}\hat \xi \left( t \right) + {\phi ^{\rm{T}}}\left( {y{\rm{,\;}}u{\rm{,\;}}\hat \delta } \right)\hat \eta \left( t \right){\rm{ + }}L\left( {y\left( t \right) - \hat y\left( t \right)} \right){\rm{,}}\hfill\\
\hat \delta \left( t \right) = h_\delta ^{\rm{T}}{\Phi _\delta }\left( t \right){{\hat x}_{\delta 0}}\left( t \right){\rm{,}}\hfill
\end{gathered}
\end{equation}
where ${A_L} = {A_0} + LC_0^{\rm{T}}$ is a Hurwitz matrix.

However, in order to implement the observer \eqref{eq35}, the estimates of ${T_I}\left( \theta  \right){\rm{,\;}}\eta \left( \theta  \right)$ and ${x_{\delta 0}}$ are required. Using regression equations \eqref{eq24} and \eqref{eq33}, we can only obtain estimates of $\eta \left( \theta  \right){\rm{,\;}}\theta $, but it is obviously not enough to implement \eqref{eq35}. Therefore, it is necessary to transform regression equations \eqref{eq24} and \eqref{eq33} into the ones w.r.t. ${T_I}\left( \theta  \right){\rm{,\;}}{x_{\delta 0}}$ such that:
\begin{equation}\label{eq36}
\begin{gathered}
{{\cal Y}_{{T_I}}}\left( t \right) = {{\cal M}_{{T_I}}}\left( t \right){T_I}\left( \theta  \right){\rm{,\;}}{{\cal M}_{{T_I}}}\left( t \right) \in \mathbb{R}{\rm{,}}\\
{{\cal Y}_{{x_{\delta 0}}}}\left( t \right) = {{\cal M}_{{x_{\delta 0}}}}\left( t \right){x_{\delta 0}}{\rm{,\;}}{{\cal M}_{{x_{\delta 0}}}}\left( t \right) \in \mathbb{R}.
\end{gathered}
\end{equation}

The first aim is to obtain the equation w.r.t. ${T_I}\left( \theta  \right)$. The matrix ${\cal P}\left( \theta  \right) \in {\mathbb{R}^{3 \times 3}}$ is formed from the denominators of the elements of ${T_I}\left( \theta  \right)$. Then the matrix ${T_I}\left( \theta  \right)$ satisfies the following regressions equation:
\begin{equation}\label{eq37}
\begin{gathered}
{\cal Q}\left( \theta  \right) = {\cal P}\left( \theta  \right){T_I}\left( \theta  \right){\rm{,}}\\
{\cal P}\left( \theta  \right) = diag\left\{ {1{\rm{,\;}}{\theta _2}{\rm{,\;}}1} \right\}{\rm{,\;}}\\
{\cal Q}\left( \theta  \right) = {\begin{bmatrix}
1&0&0\\
{ - {\theta _1}}&0&{{\theta _1}{\theta _2}{\theta _3}}\\
0&{ - {\theta _1}}&0
\end{bmatrix}},
\end{gathered}
\end{equation}
which is multiplied by 
\begin{gather*}
{\Pi _{{T_{_I}}}}\left( {{{\cal M}_\theta }\left( t \right)} \right) = {\rm{diag}}\left\{ {{{\cal M}_\theta }\left( t \right){\rm{,\;}}{\cal M}_\theta ^3\left( t \right){\rm{,\;}}{{\cal M}_\theta }\left( t \right)} \right\},    
\end{gather*}
and equation \eqref{eq33} is substituted into obtained result to write:
\begin{equation}\label{eq38}
\begin{gathered}
{{\cal T}_{\cal Q}}\left( {{{\overline \Xi }_{\cal Q}}\left( {{{\cal M}_\theta }} \right){{\cal Y}_\theta }} \right) = {{\cal T}_{\cal P}}\left( {{{\overline \Xi }_{\cal P}}\left( {{{\cal M}_\theta }} \right){{\cal Y}_\theta }} \right)\theta {\rm{,}}\\
\end{gathered}
\end{equation}
\vspace{-15pt}
\begin{gather*}
 {{\cal T}_{\cal Q}}\left( {{{\overline \Xi }_{\cal Q}}\left( {{{\cal M}_\theta }} \right){{\cal Y}_\theta }} \right) = {\begin{bmatrix}
{{{\cal M}_\theta }}&0&0\\
{ - {\cal M}_\theta ^2{{\cal Y}_{1\theta }}}&0&{{{\cal Y}_{1\theta }}{{\cal Y}_{2\theta }}{{\cal Y}_{3\theta }}}\\
0&{ - {{\cal Y}_{1\theta }}}&0
\end{bmatrix}}{\rm{,\;}}\\
{{\cal T}_{\cal P}}\left( {{{\overline \Xi }_{\cal P}}\left( {{{\cal M}_\theta }} \right){{\cal Y}_\theta }} \right) = diag\left\{ {{{\cal M}_\theta }{\rm{,\;}}{\cal M}_\theta ^2{{\cal Y}_{2\theta }}{\rm{,\;}}{{\cal M}_\theta }} \right\}{\rm{,}}   
\end{gather*}
which is then multiplied by ${\rm{adj}}\left\{ {{{\cal T}_{\cal P}}\left( {{{\overline \Xi }_{\cal P}}\left( {{{\cal M}_\theta }} \right){{\cal Y}_\theta }} \right)} \right\}$ to obtain the first equation from \eqref{eq36}:
\begin{equation}\label{eq39}
\begin{gathered}
{{\cal Y}_{{T_I}}}\left( t \right) = {{\cal M}_{{T_I}}}\left( t \right){T_I}\left( \theta  \right){\rm{,}}\\
{{\cal Y}_{{T_I}}}\left( t \right) = {\rm{adj}}\left\{ {{{\cal T}_{\cal P}}\left( {{{\overline \Xi }_{\cal P}}\left( {{{\cal M}_\theta }\left( t \right)} \right){{\cal Y}_\theta }\left( t \right)} \right)} \right\}\times\;\;\;\;\;\;\;\;\;\;\;\;\;\;\;\\\hfill
\times{{\cal T}_{\cal Q}}\left( {{{\overline \Xi }_{\cal Q}}\left( {{{\cal M}_\theta }\left( t \right)} \right){{\cal Y}_\theta }\left( t \right)} \right){\rm{,}}\\
{{\cal M}_{{T_I}}}\left( t \right) = {\rm{det}}\left\{ {{{\cal T}_{\cal P}}\left( {{{\overline \Xi }_{\cal P}}\left( {{{\cal M}_\theta }\left( t \right)} \right){{\cal Y}_\theta }\left( t \right)} \right)} \right\}{\rm{,}}
\end{gathered}
\end{equation}
where, owing to
\begin{displaymath}
\begin{gathered}
    {{\cal M}_{{T_I}}}\left( t \right) = {\rm{det}}\left\{ {{{\cal T}_{\cal P}}\left( {{{\overline \Xi }_{\cal P}}\left( {{{\cal M}_\theta }\left( t \right)} \right){{\cal Y}_\theta }\left( t \right)} \right)} \right\} =\;\;\;\;\;\;\;\;\;\;\;\;\;\;\;\;\;\;\;\;\;\;\;\;\;\\\hfill
= \underbrace {{\rm{det}}\left\{ {{\cal P}\left( \theta  \right)} \right\}}_{ > 0}\underbrace {{\rm{det}}\left\{ {{\Pi _{{T_{_I}}}}\left( {{{\cal M}_\theta }\left( t \right)} \right)} \right\}}_{ = {\cal M}_\theta ^5\left( t \right)}{\rm{,}}   
\end{gathered}
\end{displaymath}
if $\overline \varphi \left( t \right) \in {\rm{FE}}$ over $\left[ {{t_\epsilon}{\text{;\;}}{t_e}} \right]$, then for all $t \ge {t_e}$ it holds that $\left| {{{\cal M}_{{T_I}}}\left( t \right)} \right| \ge \underline {{{\cal M}_{{T_I}}}}  > 0$. 

Parametrization of the regression equation w.r.t. the initial conditions ${x_{\delta 0}}$ of the disturbance producing exosystem \eqref{eq6} is represented in the following lemma.

\textbf{Lemma 2.} \emph{The unknown parameters ${x_{\delta 0}}$ satisfy the following measurable regression equation:}
\begin{equation}\label{eq40}
\begin{gathered}
{{\cal Y}_{{x_{\delta 0}}}}\left( t \right) = {{\cal M}_{{x_{\delta 0}}}}\left( t \right){x_{\delta 0}}{\rm{,}}\\
{{{\cal Y}_{{x_{\delta 0}}}}\left( t \right) \!=\! {\rm{adj}}\left\{ {{V_f}\left( t \right)} \right\}{p_f}\left( t \right){\rm{,}}}\;
{{{\cal M}_{{x_{\delta 0}}}}\left( t \right) \!=\! {\rm{det}}\left\{ {{V_f}\left( t \right)} \right\}{\rm{,}}}
\end{gathered}
\end{equation}
\emph{where the signals ${p_f}\left( t \right)$ and ${V_f}\left( t \right)$ are obtained as follows:}
\begin{equation}\label{eq41}
\begin{array}{*{20}{c}}
  {p_f}\left( t \right) = \int\limits_{{t_\epsilon}}^t {e^{ - \sigma \left( {\tau  - {t_\epsilon}} \right)}}\Delta \left( \tau  \right){{\left( {{I_{{n_\delta }}} \otimes {\mathcal{Y}_{{\psi _d}}}\left( \tau  \right)} \right)}^{\text{T}}}{V^{\text{T}}}\left( \tau  \right)\times\\
  \times{C_0}\Delta \left( \tau  \right)p\left( \tau  \right)d\tau  {\text{, }}{p_f}\left( {{t_\epsilon}} \right) = {0_{{n_\delta }}}{\text{,}} \\ 
  p\left( t \right) = \Delta \left( t \right)\overline q\left( t \right) - C_0^{\text{T}}\Omega \left( t \right){\mathcal{L}_a}\mathcal{Y}\left( t \right) - C_0^{\text{T}}P\left( t \right){\mathcal{L}_b}\mathcal{Y}\left( t \right){\text{,}} \\ 
  {\mathcal{L}_a}\eta \left( \theta  \right) = {\psi _a}\left( \theta  \right){\text{, }}{\mathcal{L}_b}\eta \left( \theta  \right) = {\psi _b}\left( \theta  \right), \\ 
  {V_f}\left( t \right) = \int\limits_{{t_\epsilon}}^t {e^{ - \sigma \left( {\tau  - {t_\epsilon}} \right)}}{\Delta ^2}\left( \tau  \right){{\left( {{I_{{n_\delta }}} \otimes {\mathcal{Y}_{{\psi _d}}}\left( \tau  \right)} \right)}^{\text{T}}}{V^{\text{T}}}\left( \tau  \right){C_0}\times\\
  \times{C_{0}}^{\text{T}}V\left( \tau  \right)\left( {{I_{{n_\delta }}} \otimes {\mathcal{Y}_{{\psi _d}}}\left( \tau  \right)} \right)d\tau  {\text{, }}{V_f}\left( {{t_\epsilon}} \right) = {0_{{n_\delta } \times {n_\delta }}}, \\ 
  \dot V\left( t \right) = {A_K}V\left( t \right) + \left( {h_\delta ^{\text{T}}{\Phi _\delta }\left( t \right) \otimes {I_n}} \right){\text{, }}V\left( {{t_0}} \right) = {0_{n \times n{n_\delta }}}{\text{,}} \\ 
\end{array}
\end{equation}
\emph{and, if $\overline \varphi \left( t \right) \in {\rm{FE}}$ and $\left( {h_\delta ^{\rm{T}}{\Phi _\delta }\left( t \right) \otimes {I_n}} \right) \in {\rm{FE}}$ over $\left[ {{t_\epsilon}{;\;}{t_e}} \right]$, then for all $t \ge {t_e}$ it holds that $\left| {{{\cal M}_{{x_{\delta 0}}}}\left( t \right)} \right| \ge \underline {{{\cal M}_{{x_{\delta 0}}}}}  > 0$.}

\emph{Proof of Lemma 2 is given in \cite{b25, b31} up to substitution of the notation ${{\cal M}_{{\psi _d}}}\left( t \right) = \Delta \left( t \right)$ in the equation for ${p_f}\left( t \right)$.}

~

Having measurable regression equations \eqref{eq24}, \eqref{eq39} and \eqref{eq40} at hand, the following identification laws are introduced:
\begin{equation}\label{eq42}
\begin{gathered}
\dot {\hat \eta} \left( t \right) = \dot {\tilde \eta} \left( t \right) =  - \gamma \left( t \right)\Delta \left( t \right)\left( {\Delta \left( t \right)\hat \eta \left( t \right) - {\cal Y}\left( t \right)} \right){\rm{,}}\\
{{\dot {\hat x}}_{\delta 0}}\left( t \right) = {{\dot {\tilde x}}_{\delta 0}}\left( t \right) = \hfill \\ \hfill = - {\gamma _{{x_{\delta 0}}}}\left( t \right){{\cal M}_{{x_{\delta 0}}}}\left( t \right)( {{{\cal M}_{{x_{\delta 0}}}}\left( t \right){{\hat x}_{\delta 0}}\left( t \right) - {{\cal Y}_{{x_{\delta 0}}}}\left( t \right)} ){\rm{,}}\\
{{\dot {\hat T}}_I}\left( t \right) = {{\dot {\tilde T}}_I}\left( t \right) = \hfill \\ \hfill = - {\gamma _{{T_I}}}\left( t \right){{\cal M}_{{T_I}}}\left( t \right)\left( {{{\cal M}_{{T_I}}}\left( t \right){{\hat T}_I}\left( t \right) - {{\cal Y}_{{T_I}}}\left( t \right)} \right)
\end{gathered}
\end{equation}
to obtain the estimates of ${T_I}\left( \theta  \right){\rm{,\;}}\eta \left( \theta  \right){\rm{,\;}}{x_{\delta 0}}$ and, as a consequence, implement the adaptive observer \eqref{eq35}.

The conditions, under which the identification laws \eqref{eq42} ensure exponential convergence of the state observation error ${\tilde x_p}\left( t \right)$ regardless of the boundedness of the output $y\left( t \right)$ and control $u\left( t \right)$ signals, are presented in the following theorem.

\textbf{Theorem 1.} \emph{Let the estimates of ${x_p}\left( t \right)$ be obtained with the help of the adaptive observer \eqref{eq35} + \eqref{eq42}, then, if $\overline \varphi \left( t \right) \in {\rm{FE}}$, $\left( {h_\delta ^{\rm{T}}{\Phi _\delta }\left( t \right) \otimes {I_n}} \right) \in {\rm{FE}}$ and we make the following choice of adaptive gains:}
\begin{gather*}
     \gamma \left( t \right){\text{:}} = \left\{ \begin{gathered}
  0{\text{, if }}\Delta \left( t \right) < \rho  \in \left[ {0{\text{; }}{\Delta _{{\text{min}}}}} \right){\text{,}} \hfill \\
  {{{{\gamma _1} \!+\! {\gamma _0}{\lambda _{{\text{max}}}}\left( {\phi \left( {y{\text{, }}u{\text{, }}\hat \delta } \right){\phi ^{\text{T}}}\!\left( {y{\text{, }}u{\text{, }}\hat \delta } \right)} \right)}}\over{{{\Delta ^2}\left( t \right)}}}{\text{ otherwise,}} \hfill \\ 
\end{gathered}  \right.{\text{ }} \\ 
\end{gather*}
\vspace{-20pt}
\begin{equation}\label{eq43}
\begin{gathered}
  {\gamma _{{x_{\delta 0}}}}\left( t \right){\text{:}} = \left\{ \begin{gathered}
  0{\text{, if }}\Delta \left( t \right) < \rho  \in \left[ {0{\text{; }}{\Delta _{{\text{min}}}}} \right){\text{,}} \hfill \\
  \frac{{{\gamma _1}}}{{\mathcal{M}_{{x_{\delta 0}}}^2\left( t \right)}}{\text{ otherwise,}} \hfill \\ 
\end{gathered}  \right.{\text{ }}\\
  {{\gamma _1} > c},\; 
  {\left\| {{\gamma _0}\Phi _\delta ^{\text{T}}{h_\delta }\psi _d^{\text{T}}{\psi _d}h_\delta ^{\text{T}}{\Phi _\delta }} \right\| \leq c{\text{,}}}  \\  
\end{gathered}
\end{equation}
\vspace{-5pt}
\begin{gather*}
    {\gamma _{{T_I}}}\left( t \right){\text{:}} = \left\{ \begin{gathered}
  0{\text{, if }}\Delta \left( t \right) < \rho  \in \left[ {0{\text{; }}{\Delta _{{\text{min}}}}} \right){\text{,}} \hfill \\
  {{{{\gamma _1} + {\gamma _0}{\lambda _{{\text{max}}}}\left( {\dot {\hat \xi} \left( t \right){{\hat \xi }^{\text{T}}}\left( t \right) \otimes {I_3}} \right)}}\over{{{\mathcal{M}_{{T_I}}^2\left( t \right)}}}}{\text{ otherwise,}} \hfill \\ 
\end{gathered}\right.
\end{gather*}
\emph{then, regardless of the boundedness of $u\left( t \right)$ and $y\left( t \right)$, it holds for the augmented observation error \linebreak ${\zeta _o} = {{\begin{bmatrix}
{\tilde x_p^{\rm{T}}\left( t \right)}&{{{\tilde \eta }^{\rm{T}}}\left( t \right)}&{\tilde x_{\delta 0}^{\rm{T}}\left( t \right)}&{ve{c^{\rm{T}}}\left\{ {{{\tilde T}_I}\left( t \right)} \right\}}
\end{bmatrix}}^{\rm{T}}}$ that: }
\begin{enumerate}
    \item[1)] ${\zeta _o}\left( t \right) \in {L_\infty }{\rm{\;\;}}\forall t \in \left[ {{t_0}{\rm{;\;}}{t_e}} \right){\rm{;}}$
    \item[2)] $\forall t \ge {t_e}\;\;\mathop {{\rm{ lim}}}\limits_{t \to \infty } \left\| {{\zeta _o}\left( t \right)} \right\| = 0{\rm{ }}\left( {{\rm{exp}}} \right).$
\end{enumerate}

\emph{Proof of Theorem 1 is given in Appendix.}

~

Now the estimates of the system \eqref{eq4} states are available, and, therefore, the only step we need to take to implement the control law \eqref{eq18} is to obtain the estimates of $\kappa \left( \theta  \right)$.

\textbf{Remark 1.} \emph{The classic adaptive state observers \cite{b13, b14, b15} usually require {\it a priori} control/output signals boundedness, which restricts their applicability to control problems and does not allow one to apply the "separation principle" well-known in the linear control theory. Owing to a special choice \eqref{eq43} of the adaptive gains of the identification laws \eqref{eq42}, the proposed observer \eqref{eq35} + \eqref{eq42} ensures exponential convergence to zero of the state observation error regardless of the control/output signals boundedness. The interpretation of this feature is as follows: regardless of the growth rate of the function $\phi \left( {y{\rm{,\;}}u{\rm{,\;}}\delta } \right)$, if adaptive gains are chosen as \eqref{eq43}, then it holds for the multiplication ${\phi ^{\rm{T}}}\left( {y{\rm{,\;}}u{\rm{,\;}}\delta } \right)\tilde \eta \left( t \right)$ in \eqref{eqA3} that ${\phi ^{\rm{T}}}\left( {y{\rm{,\;}}u{\rm{,\;}}\delta } \right)\tilde \eta \left( t \right) \to 0{\rm{ }}\left( {{\rm{exp}}} \right)$ (for instance, $\phi \left( {y{\rm{,\;}}u{\rm{,\;}}\delta } \right) = {e^t}$,  $\tilde \eta \left( t \right) = {e^{ - 2t}}$), which, owing to the fact that ${A_L}$ is a Hurwitz matrix, is necessary and sufficient to ensure exponential convergence of ${\tilde x_p}\left( t \right)$. The fact that the error ${\tilde x_p}\left( t \right)$ is bounded over the time range $\left[ {{t_0}{\text{; }}{t_\epsilon}} \right)$ follows from the Theorem of existence and uniqueness of solution of differential equations for ${x_p}\left( t \right)$ and $\xi \left( t \right)$.}

\subsection{Controller Parameters Identification}

Similarly to reasoning under equations \eqref{eq13} and \eqref{eq16}, we use Assumption 3 to write \eqref{eq14}, which is then multiplied by ${\rm{adj}}\left\{ {{{\cal T}_{\cal G}}\left( {{{\overline \Xi }_{\cal G}}\left( \Delta  \right){{\cal Y}_\theta }} \right)} \right\}$ to obtain the regression equation w.r.t. $\kappa \left( \theta  \right)$ with the scalar regressor, which is bounded away from zero for all $t \ge {t_e}$:
\begin{equation}\label{eq44}
\begin{gathered}
{{\cal Y}_\kappa }\left( t \right) = {{\cal M}_\kappa }\left( t \right)\kappa \left( \theta  \right),\\
{{\cal Y}_\kappa }\left( t \right) = {\rm{adj}}\left\{ {{{\cal T}_{\cal G}}\left( {{{\overline \Xi }_{\cal G}}\left( \Delta  \right){{\cal Y}_\theta }} \right)} \right\}{{\cal T}_{\cal S}}\left( {{{\overline \Xi }_{\cal S}}\left( \Delta  \right){{\cal Y}_\theta }} \right){\rm{,\;}}\\
{{\cal M}_\kappa }\left( t \right) = {\rm{det}}\left\{ {{{\cal T}_{\cal G}}\left( {{{\overline \Xi }_{\cal G}}\left( \Delta  \right){{\cal Y}_\theta }} \right)} \right\}{\rm{,\;}}
\end{gathered}
\end{equation}
where, if $\overline \varphi \left( t \right) \in {\rm{FE}}$ over $\left[ {{t_\epsilon}{;\;}{t_e}} \right]$, then for all $t \ge {t_e}$ it holds that $\left| {{{\cal M}_\kappa }\left( t \right)} \right| \ge \underline {{{\cal M}_\kappa }}  > 0$. 

Using \eqref{eq44}, the identification law of ideal controller \eqref{eq10} parameters is introduced:
\begin{equation}\label{eq45}
\begin{gathered}
\dot {\hat \kappa} \left( t \right) \!=\! \dot {\tilde \kappa} \left( t \right) \!=\!  - {\gamma _\kappa }\left( t \right){{\cal M}_\kappa }\left( t \right)\left( {{{\cal M}_\kappa }\left( t \right)\hat \kappa \left( t \right) - {{\cal Y}_\kappa }\left( t \right)} \right).
\end{gathered}
\end{equation}

The conditions, under which the control law \eqref{eq18} with observer \eqref{eq35} and identification laws \eqref{eq42}, \eqref{eq45} ensure that the goal \eqref{eq19} is achieved, are given in the following theorem.

\textbf{Theorem 2.} \emph{Let the states ${x_p}\left( t \right)$ estimates be obtained with the help of the adaptive observer \eqref{eq35} + \eqref{eq42} + \eqref{eq43}, and the estimates of $\kappa \left( \theta  \right)$ be formed by the identification laws \eqref{eq45}, and the adaptive gains be chosen as:}
\begin{equation}\label{eq46}
\begin{gathered}
{\gamma _\kappa }\left( t \right){\text{:}} = \left\{ \begin{gathered}
  0{\text{, if }}\Delta \left( t \right) < \rho  \in \left[ {0{\text{; }}{\Delta _{{\text{min}}}}} \right){\text{,}} \hfill \\
  \frac{{{\gamma _1} + {\gamma _0}{\lambda _{{\text{max}}}}\left( {\hat x\left( t \right){{\hat x}^{\text{T}}}\left( t \right)} \right)}}{{\mathcal{M}_\kappa ^2\left( t \right)}}{\text{ otherwise,}} \hfill \\ 
\end{gathered}  \right.
\end{gathered}
\end{equation}
\emph{then the control law \eqref{eq18} ensures for the augmented error $\zeta \left( t \right)$ that:}
\begin{enumerate}
    \item [1)] $\zeta \left( t \right) \in {L_\infty }{\rm{\;\;}}\forall t \in \left[ {{t_0}{\rm{;\;}}{t_e}} \right){\rm{;}}$
    \item [2)] $\forall t \ge {t_e}\mathop {{\rm{lim}}}\limits_{t \to \infty } \left\| {\zeta \left( t \right)} \right\| = 0{\rm{ }}\left( {{\rm{exp}}} \right).$
\end{enumerate}

\emph{Proof of Theorem 2 is presented in Appendix.}

~

A block diagram of the proposed system of adaptive vibration suppression is presented in Figure 1.

\begin{figure}[!thpb]
\centering
\includegraphics[width=3.5in]{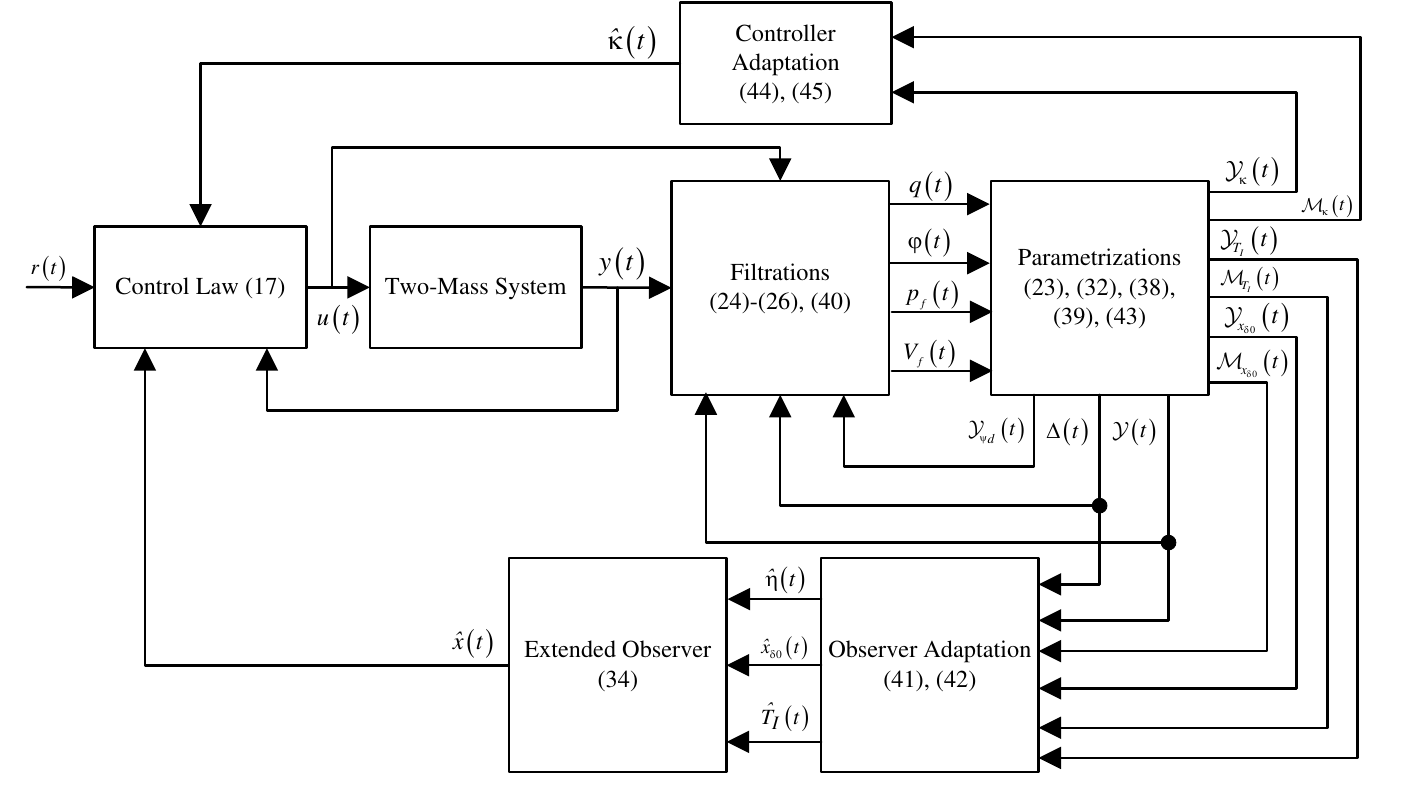}
\caption{Block diagram of proposed sensorless adaptive vibration suppression control system}
\label{fig_1}
\end{figure}

As follows from Fig. 1, the proposed control system consists of measured signals filtering \eqref{eq25}-\eqref{eq27}, \eqref{eq41}, parameterization schemes of regression equations w.r.t. parameters of the full-information-based control law \eqref{eq18} and observer \eqref{eq35}, and corresponding adaptive laws \eqref{eq45}, \eqref{eq46} and \eqref{eq42}, \eqref{eq43}. Compared to the existing schemes of adaptive vibration suppression for two-mass systems from \cite{b12, b16, b17, b18, b19, b21, b22, b23}, the proposed solution: ({\it i}) does not use singularity causing operation of division by the dynamic estimates of the unknown parameters, ({\it ii}) rigorously defines the conditions of exponential stability of the closed-loop system, ({\it iii}) ensures exponential stability of the closed-loop adaptive control system when the regressor is finitely exciting. 

To the best of the authors’ knowledge, the proposed adaptive control structure is a first mathematically sound solution of adaptive vibration suppression for two-mass systems.

\textbf{Remark 2.} \emph{The requirements on how to choose ${\gamma _{\left( . \right)}}$ from Theorems 1 and 2 are conservative with degree defined by the upper bounds used in the analysis of the Lyapunov candidate function derivative. Numerical experiments show that the identification laws can be applied with time-invariant adaptive gains, which means that we do not need to implement the functionals ${\lambda _{{\rm{max}}}}\left( . \right)$ in laws \eqref{eq42}, \eqref{eq45} and choose the parameter $\rho $. At the same time, Theorems 1 and 2 do not prohibit to choose the parameter ${\gamma _0}$ from the condition ${\gamma _0} > 0{\rm{,\;}}{\gamma _0} \to 0$ , which, in fact, allows one to avoid implementation of the functionals ${\lambda _{{\rm{max}}}}\left( . \right)$ in practice and, at the same time, not to violate the premises of Theorems.}

\subsection{Ad Hoc Practical Oriented Modifications}

It should be noted that the integrand of filters \eqref{eq25} and \eqref{eq41} exponentially decreases for all $t \ge {t_{\epsilon}}$. Consequently, when $t$ is sufficiently large, the proposed adaptive control system loses its awareness and is not able to track changes of the parameters $\theta $ in case they are not time-invariant, and, as a result, suppress the vibration in the two-mass system.

To overcome this drawback, the states of filters \eqref{eq25} and \eqref{eq41} should be reset to zero:
\begin{equation}\label{eq47}
\begin{array}{*{20}{c}}
q\left( t \right) = \int\limits_{t_i^ + }^t {e^{ - \sigma \left( {\tau  - t_i^ + } \right)}}{{\overline \varphi }_f}\left( \tau  \right)( \overline q\left( \tau  \right) - {k_1}{{\overline q}_f}\left( \tau  \right) -\hfill\\\hfill
-{\beta ^{\rm{T}}}\left( {{F_f}\left( \tau  \right) + l{y_f}\left( \tau  \right)} \right))d\tau {\rm{,\;}}\\
\varphi \left( t \right) = \int\limits_{t_i^ + }^t {{e^{ - \sigma \left( {\tau  - t_i^ + } \right)}}{{\overline \varphi }_f}\left( \tau  \right)\overline \varphi _f^{\rm{T}}\left( \tau  \right)d\tau } {\rm{,\;}}\\
q\left( {t_i^ + } \right) = {0_{{\rm{2}}n}},\;\varphi \left( {t_i^ + } \right) = {0_{{\rm{2}}n \times {\rm{2}}n}},
\end{array}
\end{equation}

\begin{equation}\label{eq48}
\begin{array}{*{20}{c}}
  {p_f}\left( t \right) = \int\limits_{t_i^ + }^t {e^{ - \sigma \left( {\tau  - t_i^ + } \right)}}{{\left( {{I_{{n_\delta }}} \otimes {\mathcal{Y}_{{\psi _d}}}\left( \tau  \right)} \right)}^{\text{T}}}{V^{\text{T}}}\left( \tau  \right)\times\hfill\\\hfill
  \times{C_0}{\Delta ^2}\left( \tau  \right)p\left( \tau  \right)d\tau {\text{, }} \\ 
  {V_f}\left( t \right) = \int\limits_{t_i^ + }^t {e^{ - \sigma \left( {\tau  - t_i^ + } \right)}}{{\left( {{I_{{n_\delta }}} \otimes {\mathcal{Y}_{{\psi _d}}}\left( \tau  \right)} \right)}^{\text{T}}}{V^{\text{T}}}\left( \tau  \right){C_0}\times\hfill\\\hfill
  \times{C_{0}}^{\text{T}}V\left( \tau  \right)\left( {{I_{{n_\delta }}} \otimes {\mathcal{Y}_{{\psi _d}}}\left( \tau  \right)} \right){\Delta ^2}\left( \tau  \right)d\tau {\text{, }} \\ 
  {p_f}\left( {t_i^ + } \right) = {0_{{n_\delta }}}, \; {V_f}\left( {t_i^ + } \right) = {0_{{n_\delta } \times {n_\delta }}}{\text{,}}
\end{array}
\end{equation}
where $t_i^ +  \in \Im  = \left\{ {\left. {t_0^ + {\text{, }}t_1^ + {\text{, }}{\kern 1pt}  \ldots {\text{, }}t_{i - 1}^ +  \ldots } \right|i \in \mathbb{N}} \right\}$ and $t_0^ +  = {t_\epsilon}$.

The elements of the sequence $\Im $ can be defined in different ways. In \cite{b32}, using a special detection scheme, it is proposed to reset the filter states to zero at time instants of the system parameters switch. In \cite{b33} it is proposed to reset the filters when the reference value is changed.

In practice, the controller \eqref{eq18} and observer \eqref{eq35} parameters adjustment (and states reset for filters \eqref{eq25} and \eqref{eq41}) is required in case of significant deterioration of the transients quality. Therefore, for practical scenarios, it is reasonable to choose the sequence $\Im $ elements in accordance with some rule:
\begin{equation}\label{eq49}
\begin{gathered}
{\rm{IF }}f\left( {y{\rm{,\;}}\hat y{\rm{,\;}}t} \right) = {f_{{\rm{max}}}}{\rm{\;THEN\;}}t_i^ +  = t{\rm{,}} 
\end{gathered}
\end{equation}
where ${f_{{\rm{max}}}} > 0$ is the acceptable upper limit of the quality criterion $f\left( {y{\rm{,\;}}\hat y{\rm{,\;}}t} \right) \ge 0{\rm{,\;\;}}f\left( {y{\rm{,\;}}\hat y{\rm{,\;}}t_i^ + } \right) = 0$, which satisfies the monotonicity condition:
\begin{displaymath}
f\left( {y{\rm{,\;}}\hat y{\rm{,\;}}{t_a}} \right) \ge f\left( {y{\rm{,\;}}\hat y{\rm{,\;}}{t_b}} \right){\rm{,\;}}\forall {t_a} \ge {t_b}{\rm{\;,\;}}{t_a}{\rm{,\;}}{t_b} \in \left[ {t_{i - 1}^ + {\rm{;\;}}t_i^ + } \right){\rm{.}}    
\end{displaymath}

The reference is usually a piecewise-constant signal, which is insufficient to meet the condition $\overline \varphi \left( t \right) \in {\rm{FE}}$ \cite{b34}. Therefore, after each reset of filters \eqref{eq47} and \eqref{eq48} it is necessary to inject a vanishing periodic component to a main-line reference signal, which ensures that the condition $\overline \varphi \left( t \right) \in {\rm{FE}}$ of exponential stability of the error $\zeta \left( t \right)$ is met:
\begin{equation}\label{eq50}
\begin{gathered}
r\left( t \right) = c\left( t \right) + d\left( t \right) =\;\;\;\;\;\;\;\;\;\;\;\;\;\;\;\;\;\;\;\;\;\;\;\;\;\;\;\;\;\;\;\;\;\;\;\;\;\;\;\;\;\;\;\;\;\;\\\;\;\;\;\;\;\;\;\;\;\;
= c\left( t \right) + h\left( {t - t_i^ + } \right){e^{ - \alpha \left( {t - t_i^ + } \right)}}\sum\limits_{j = 0}^{{n_d}} {{a_j}{\rm{sin}}\left( {{\omega _j}t} \right)} {\rm{,}}\\
{\omega _j} \ne {\omega _k}{\rm{,\;}}\forall j \ne k{\rm{,\;}}{a_j} > 0{\rm{,\;}}\alpha  > 0{\rm{,\;}}{n_d} \ge 1{\rm{,}}
\end{gathered}
\end{equation}
where $c\left( t \right)$ is a main-line piecewise-constant reference, $d\left( t \right)$ is a dither signal to ensure that FE condition is met, $h\left( {t - t_i^ + } \right)$ is the Heaviside step function.

\section{Numerical experiments}

The proposed adaptive control system has been implemented in Matlab/Simulink environment. The load torque and parameters of system \eqref{eq4} were set (in per unit) as:
\begin{equation}\label{eq51}
\begin{gathered}
{\theta _1} = 0.203{\rm{,\;}}{\theta _2} = \left\{ \begin{array}{l}
{\theta _1}{\rm{,\;if\;}}t \in \left[ {0{\rm{;\;11}}} \right)\\
1.75{\theta _1}{\rm{,\;if\;}}t \ge 11
\end{array} \right.{\rm{,}}\\
{\theta _3} = 0.0026{\rm{,\;}}\delta  = \left\{ \begin{array}{l}
0.5{\rm{,\;if\;}}t \in \left[ {0{\rm{;\;27}}} \right)\\
 - 0.5{\rm{,\;if\;}}t \ge 27
\end{array} \right.{\rm{,}}\\
{h_\delta } = 1{\rm{,\;}}{{\cal A}_\delta } = 0.
\end{gathered}
\end{equation}

A pole-placement-based control law defined by \eqref{eq15}, \eqref{eq16} was considered as a full-information-based ideal control law \eqref{eq10}. The parameters of filters \eqref{eq26}, \eqref{eq27}, \eqref{eq47}, \eqref{eq48} and equation \eqref{eq16} to calculate the coefficients of the control law \eqref{eq18} were picked as:
\begin{equation}\label{eq52}
\begin{gathered}
  \sigma \!\left( {{A_K}} \right) \!=\! \sigma \!\left( {{A_L}} \right) \!=\!  - 50{I_3}{\text{, }}l \!=\! 5{\text{, }}G =  - 5{\text{, }}\beta  \!=\! {M_\delta } \!=\! 1{\text{, }} \\ 
  {k_1} = 25{\text{, }}\sigma  = 5{\text{, }}k\left( t \right) = {10^{63}}{\text{, }}{\gamma _1} = 1{\text{, }}{\gamma _0} = {10^{ - 11}}{\text{,}} \\ 
  {\xi _d} = 0.7{\text{, }}{\omega _d} = 25{\text{, }}\hat \kappa \left( 0 \right) = {\begin{bmatrix}
  {60}&{ - 1.5}&0&0 
\end{bmatrix}} {\text{,}} \\ 
  {{\hat T}_I}\left( 0 \right) = {0_{3 \times 3}}{\text{, }}\hat \eta \left( 0 \right) = {0_9}{\text{, }}{{\hat x}_{\delta 0}}\left( 0 \right) = 0{\text{, }}t_0^ +  = {t_\epsilon}. \\ 
\end{gathered}
\end{equation}

The main-line reference $c\left( t \right)$ was formed as a sequence of unit square waves (see Figure 2). The parameters of the dither signal $d\left( t \right)$ were set as: 
\begin{equation}\label{eq53}
\begin{gathered}
{n_d} = 1{\rm{,\;}}\alpha  = 1{\rm{,\;}}{a_1} = 0.15{\rm{,\;}}{\omega _1} = 10.
\end{gathered}
\end{equation}

The filters with resetting \eqref{eq47}, \eqref{eq48} were applied, and the following transient quality criterion was chosen to determine resetting time instants $t_i^ + $:
\begin{equation}\label{eq54}
\begin{gathered}
f\left( {y{\rm{,\;}}\hat y{\rm{,\;}}t} \right) = \int\limits_{t_i^ + }^t {{{\left( {y\left( \tau  \right) - \hat y\left( \tau  \right)} \right)}^2}d\tau } {\rm{,\;\;}}{f_{{\rm{max}}}} = 0.01.
\end{gathered}
\end{equation}

Figure 2 presents transients of $c\left( t \right)$ and $y\left( t \right)$ for \linebreak $\hat \kappa \left( 0 \right) = {\begin{bmatrix}
{60}&{ - 1.5}&0&0
\end{bmatrix}}$ in case the identification laws \eqref{eq42} and \eqref{eq45} were disabled.
\begin{figure}[!thpb]
\centering
\includegraphics[width=3.3in]{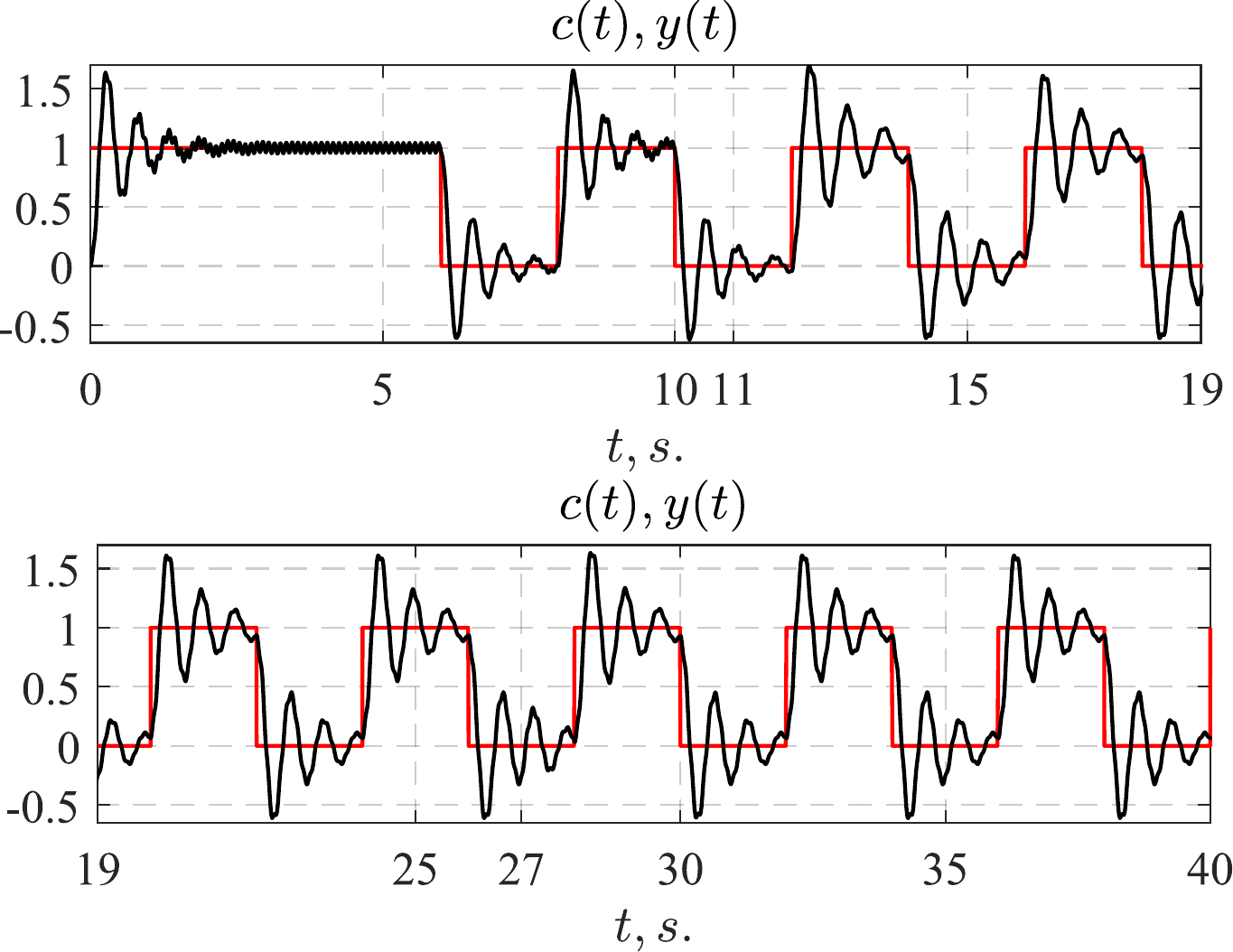}
\caption{Transients of $c\left( t \right)$ and $y\left( t \right)$ when identification laws \eqref{eq42} and \eqref{eq45} were disabled}
\label{fig_2}
\end{figure}

The obtained transients demonstrate significant undamped oscillations of the output signal when non-adaptive control system with parameters ${\rm{ }}\hat \kappa \left( 0 \right)$ was used to control the two-mass system.

Figure 3 depicts the transients of $r\left( t \right)$ and $y\left( t \right)$ in case the proposed adaptive control system was applied.
\begin{figure}[!thpb]
\centering
\includegraphics[width=3.3in]{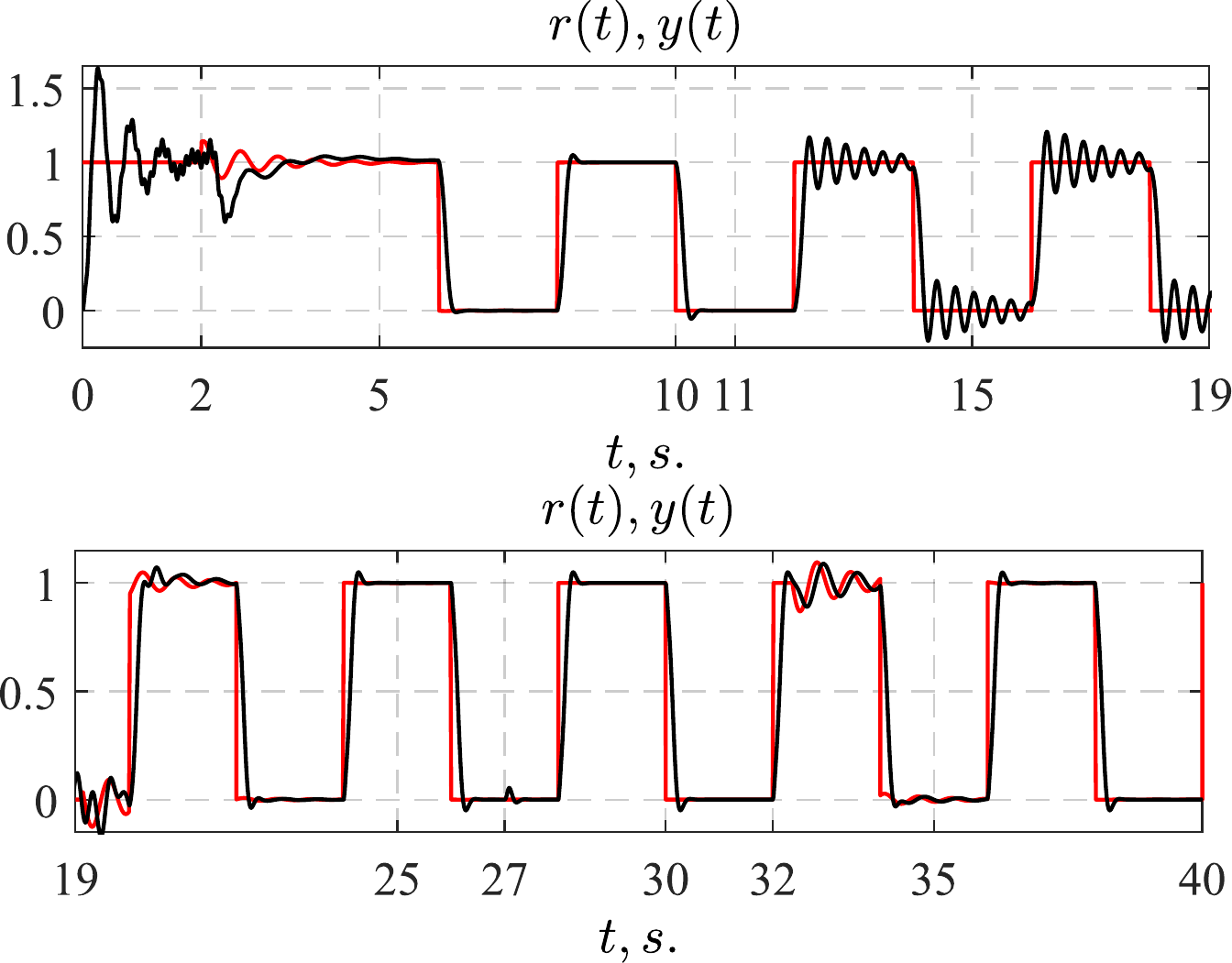}
\caption{Transients of $r\left( t \right)$ and $y\left( t \right)$ when identification \eqref{eq42} and \eqref{eq45} were applied}
\label{fig_3}
\end{figure}

Thus, the application of the developed adaptive system made it possible to suppress the vibrations of the two-mass system and significantly improved the quality of the transients. The interpretation of the obtained results is as follows. The time range $\left[ {0{\rm{;\;2}}} \right)$ was used to wait when the disturbance in the parameterization vanished sufficiently (see the proof of Lemma 1). At the time instant $t = 2$ a dither signal $d\left( t \right)$ was added to the main-line reference $c\left( t \right)$ to meet the condition $\overline \varphi \left( t \right) \in {\rm{FE}}$, and the process of the controller/observer parameters adjustment was started. Upon completion of the adaptation, as can be seen from Figure 3, the transients quality improved significantly. A load inertia moment ${\theta _2}$ change occurred at $t = 11$ and caused undamped vibrations. According to the chosen criterion \eqref{eq54}, at $t \approx 19$ the transient quality degradation was detected, a dither signal was added to the main-line reference {\it a novo} and the controller/observer parameters were readjusted. At the time instant $t = 27$ the load torque $\delta $ change occurred. According to the chosen criterion \eqref{eq54}, at $t \approx 32$ the transient quality degradation (caused by the system parameters change) was detected, a dither signal was added to the main-line reference {\it a novo}, and the controller/observer parameters were readjusted. It should be noted that the inertia moment ${\theta _2}$ change affected the transients quality of both ${e_P}\left( t \right){\rm{,\;}}y\left( t \right) - \hat y\left( t \right)$, while the load torque $\delta $ change affected only $y\left( t \right) - \hat y\left( t \right)$.

Figure 4 shows the transients of the normalized parametric errors and the norm of the state observation error.
\begin{figure}[!thpb]
\centering
\includegraphics[width=3.3in]{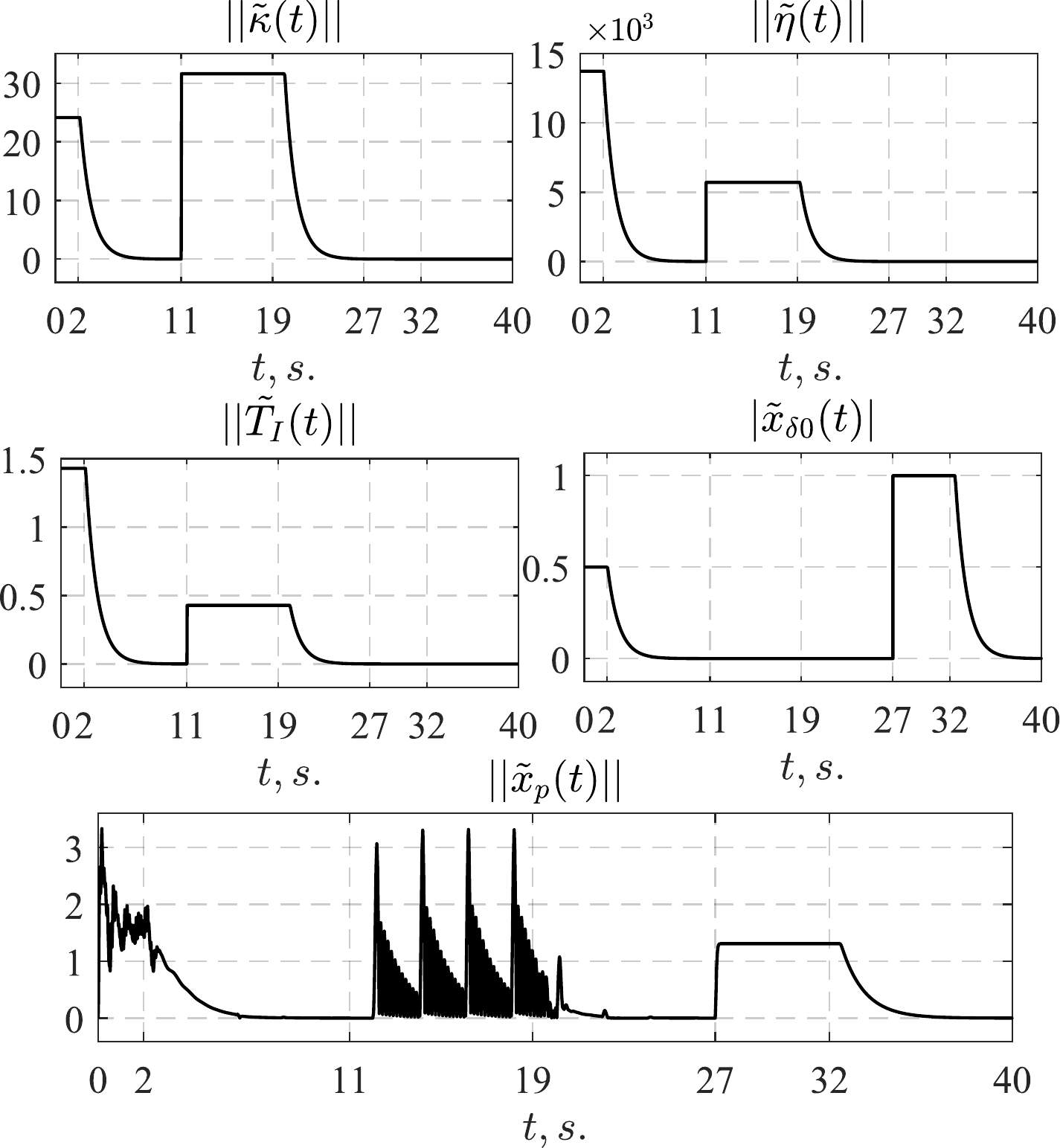}
\caption{Transients of parametric and state observation errors}
\label{fig_4}
\end{figure}

The results presented in Figure 4 confirm the convergence of the parameter/state estimates formed by the identification laws \eqref{eq42}, \eqref{eq45} and observer \eqref{eq35}, respectively, to their true values if the finite excitation condition $\overline \varphi \left( t \right) \in {\rm{FE}}$ is met. On the whole, all the simulation results are fully in line with the theoretical conclusions made in lemmas and theorems.

\section{Conclusion}

A new system of adaptive sensorless vibration suppression for two-mass electromechanical systems was proposed. If the regressor was finitely exciting, the solution ensured exponential convergence of both the identification error of ideal controller parameters (coefficients of PI-controller and additional feedbacks) and physical states observation error (load speed, load torque, torsional torque, motor speed). Further the proposed adaptive control structure can be used for sensorless adaptive control of a wide class of dynamic systems.

\renewcommand{\theequation}{A\arabic{equation}}
\setcounter{equation}{0}  

{\appendices
\section{Proof of Lemma 1}
According to the results of Lemma 1 from \cite{b25, b31}, using filtration \eqref{eq25}-\eqref{eq27}, for all $t \ge {t_e}$ we have the measurable regression equation: 
\begin{equation}\label{eqA1}
\begin{gathered}
k\left( t \right) \cdot {\rm{adj}}\left\{ {\varphi \left( t \right)} \right\}q\left( t \right) = \Delta \left( t \right){\begin{bmatrix}
{{\psi _a}\left( \theta  \right)}\\
{{\psi _b}\left( \theta  \right)}
\end{bmatrix}}{\rm{,}}
\end{gathered}
\end{equation}
then, to prove Lemma 1, \eqref{eqA1} is to be augmented with the equation w.r.t. ${\psi _d}\left( \theta  \right)$.

Following the definition \eqref{eq23}, it holds that:
\begin{displaymath}
\begin{gathered}
{\psi _d}\left( \theta  \right) = {\begin{bmatrix}
0\\
0\\
{{\textstyle{{ - 1} \over {{\theta _1}{\theta _2}{\theta _3}}}}}
\end{bmatrix}} = {\rm{ }}{\begin{bmatrix}
0&0&0\\
0&0&0\\
0&0&{ - 1}
\end{bmatrix}}{\psi _b}\left( \theta  \right) =\\
={\begin{bmatrix}
0&0&0\\
0&0&0\\
0&0&{ - 1}
\end{bmatrix}}{\begin{bmatrix}
{\theta _1^{ - 1}}\\
0\\
{{\textstyle{1 \over {{\theta _1}{\theta _2}{\theta _3}}}}}
\end{bmatrix}}{\rm{,}}
\end{gathered}
\end{displaymath}
and, as a result, owing to $\Delta \left( t \right) \in \mathbb{R}$, the equation for ${\psi _d}\left( \theta  \right)$ is rewritten as:
\begin{equation}\label{eqA2}
\begin{gathered}
k\left( t \right) \cdot {{\cal L}_d}{\rm{adj}}\left\{ {\varphi \left( t \right)} \right\}q\left( t \right) =\hfill\\\;\;\;\;\;\;\;\;\;\;\;\;\;\;\;\;\;\;\;\;\;\;\;\;\;\;
= \Delta \left( t \right){{\cal L}_d}{\begin{bmatrix}
{{\psi _a}\left( \theta  \right)}\\
{{\psi _b}\left( \theta  \right)}
\end{bmatrix}} = \Delta \left( t \right){\psi _d}\left( \theta  \right){\rm{,}}
\end{gathered}
\end{equation}
which allows one to obtain the desired equation \eqref{eq24} by combining \eqref{eqA1} and \eqref{eqA2}.

\section{Proof of Theorem 1}
Proof is divided into two steps. The first one is to show that ${\zeta _o}\left( t \right)$ is bounded for all $t \in \left[ {{t_0}{\rm{,\;}}{t_e}} \right)$, and the second one is to prove exponential convergence of ${\zeta _o}\left( t \right)$ for all $t \ge {t_e}$.

\textbf{Step 1.} State observation error equations for $\tilde \xi \left( t \right)$ and ${\tilde x_p}\left( t \right)$ are written as:	
\begin{equation}\label{eqA3}
\begin{gathered}
\dot {\tilde \xi} \left( t \right) \!=\! {A_0}\hat \xi \left( t \right) \!+\! {\phi ^{\rm{T}}}\left( {y{\rm{,\;}}u{\rm{,\;}}\hat \delta } \right)\hat \eta \left( t \right){\rm{ \!+\! }}L\left( {y\left( t \right) \!-\! \hat y\left( t \right)} \right) \!-\\
- {A_0}\xi \left( t \right) - {\phi ^{\rm{T}}}\left( {y{\rm{,\;}}u{\rm{,\;}}\delta } \right)\eta \left( \theta  \right){\rm{ = }}\hfill\\
 = {A_L}\tilde \xi \left( t \right) + {\phi ^{\rm{T}}}\left( {y{\rm{,\;}}u{\rm{,\;}}\hat \delta } \right)\hat \eta \left( t \right) -\hfill\\
 \hfill- {\phi ^{\rm{T}}}\left( {y{\rm{,\;}}u{\rm{,\;}}\delta } \right)\eta \left( \theta  \right) \pm {\phi ^{\rm{T}}}\left( {y{\rm{,\;}}u{\rm{,\;}}\hat \delta } \right)\eta \left( \theta  \right) = \\
 = {A_L}\tilde \xi \left( t \right) + {\phi ^{\rm{T}}}\left( {y{\rm{,\;}}u{\rm{,\;}}\hat \delta } \right)\tilde \eta \left( t \right) -\hfill\\
 \hfill - {\phi ^{\rm{T}}}\left( {y{\rm{,\;}}u{\rm{,\;}}\delta } \right)\eta \left( \theta  \right) + {\phi ^{\rm{T}}}\left( {y{\rm{,\;}}u{\rm{,\;}}\hat \delta } \right)\eta \left( \theta  \right) = \\
 = {A_L}\tilde \xi \left( t \right) + {\phi ^{\rm{T}}}\left( {y{\rm{,\;}}u{\rm{,\;}}\hat \delta } \right)\tilde \eta \left( t \right) + {\psi _d}\left( \theta  \right)\tilde \delta \left( t \right) =\hfill \\
 =\! {A_L}\tilde \xi \left( t \right) \!+\! {\phi ^{\rm{T}}}\!\left( {y{\rm{,\;}}u{\rm{,\;}}\hat \delta } \right)\tilde \eta \left( t \right) \!+\! {\psi _d}\left( \theta  \right)\!h_\delta ^{\rm{T}}{\Phi _\delta }\!\left( t \right){{\tilde x}_{\delta 0}}\left( t \right){\rm{,}}
\end{gathered}
\end{equation}
\begin{equation}\label{eqA4}
\begin{gathered}
{\tilde x_p}\left( t \right) = {\hat T_I}\left( t \right)\hat \xi \left( t \right) - {T_I}\left( \theta  \right)\xi \left( t \right) \pm {T_I}\left( \theta  \right)\hat \xi \left( t \right) =\hfill\\
= {\tilde T_I}\left( t \right)\hat \xi \left( t \right) + {T_I}\left( \theta  \right)\tilde \xi \left( t \right).\hfill
\end{gathered}
\end{equation}

In conservative case for all $t \in \left[ {{t_0}{\rm{,\;}}{t_e}} \right)$ it holds that \linebreak $\Delta \left( t \right) < \rho  \Rightarrow \gamma \left( t \right) = {\rm{0,\;}}{\gamma _{{x_{\delta 0}}}}\left( t \right) = {\rm{0,\;}}{\gamma _{{T_I}}}\left( t \right) = {\rm{0}}$ (it is assumed that $\Delta \left( t \right) = \rho  \Rightarrow t = {t_e}$), and, therefore, identification procedure is not active ${\dot {\tilde T}_I}\left( t \right) = 0{\rm{,\;}}\dot {\tilde \eta} \left( t \right) = 0{\rm{,\;}}{\dot {\tilde x}_{\delta 0}}\left( t \right) = 0$ over the time range $\left[ {{t_0}{\rm{,\;}}{t_e}} \right)$, and the following inequalities hold:
\begin{equation}\label{eqA5}
\begin{gathered}
{\tilde T_I}\left( t \right) = {\tilde T_I}\left( {{t_0}} \right) < \infty {\rm{,\;}}\tilde \eta \left( t \right) = \tilde \eta \left( {{t_0}} \right) < \infty {\rm{,\;}}\\
{\tilde x_{\delta 0}}\left( t \right) = {\tilde x_{\delta 0}}\left( {{t_0}} \right) < \infty .
\end{gathered}
\end{equation}

It follows from the theorem of existence and uniqueness of solution of differential equations (Theorem 3.2 from \cite{b35}) that the observation error $\tilde \xi \left( t \right)$ is bounded over $\left[ {{t_0}{\rm{,\;}}{t_e}} \right)$. Consequently, owing to ${\tilde T_I}\left( t \right) = {\tilde T_I}\left( {{t_0}} \right) < \infty $, it is concluded that ${\tilde x_p}\left( t \right)$ is also bounded over $\left[ {{t_0}{\rm{,\;}}{t_e}} \right)$, which, taking \eqref{eqA5} into consideration, means that the augmented error ${\zeta _o}\left( t \right)$ is bounded over $\left[ {{t_0}{\rm{,\;}}{t_e}} \right)$.

\textbf{Step 2.} The first aim is to prove that the error $\tilde \xi \left( t \right)$ converges exponentially to zero. To this end, a Lyapunov function candidate is introduced:
\begin{equation}\label{eqA6}
\begin{gathered}
{V_{\tilde \xi }} = {\gamma _0}{\tilde \xi ^{\rm{T}}}P\tilde \xi  + {\tilde \eta ^{\rm{T}}}\tilde \eta  + \tilde x_{\delta 0}^{\rm{T}}{\tilde x_{\delta 0}}{\rm{,\;}}
\end{gathered}
\end{equation}
where $P$ is a solution of the following set of equations under conditions $K = {k^2}{I_{n \times n}}{\rm{,\;}}D = 0.5{k^2}{I_{n \times n}}{\rm{,\;}}k = 1{\rm{,\;}}B = {I_{n \times n}}$:
\begin{displaymath}
\begin{gathered}
A_L^{\rm{T}}P + P{A_L} =  - Q{Q^{\rm{T}}} - \mu P{\rm{,\;}}P{I_{{\mathop{\rm n}\nolimits}  \times n}} = QK{\rm{,\;}}\\
{K^{\rm{T}}}K = D + {D^{\rm{T}}},        
\end{gathered}
\end{displaymath}
which is equivalent to the Riccatti equation with a bias term:
\begin{displaymath}
\begin{gathered}
A_L^{\rm{T}}P + P{A_L} + P{P^{\rm{T}}} + \mu P = {0_{n \times n}}.
\end{gathered}
\end{displaymath}

Owing to \eqref{eqA3}, \eqref{eq42}, \eqref{eq43}, the derivative of the quadratic form \eqref{eqA6} is written as:
\begin{equation}\label{eqA7}
\begin{gathered}
{{\dot V}_{\tilde \xi }} = {\gamma _0}\left[ {{\tilde \xi }^{\rm{T}}}\left( {A_L^{\rm{T}}P + P{A_L}} \right)\tilde \xi  + {{\tilde \xi }^{\rm{T}}}P{{\hat \phi }^{\rm{T}}}\tilde \eta  + {{\tilde \eta }^{\rm{T}}}\hat \phi P\tilde \xi  +\right.\\\hfill
\left.+\tilde x_{\delta 0}^{\rm{T}}\Phi _\delta ^{\rm{T}}{h_\delta }\psi _d^{\rm{T}}P\tilde \xi  + {{\tilde \xi }^{\rm{T}}}P{\psi _d}h_\delta ^{\rm{T}}{\Phi _\delta }{{\tilde x}_{\delta 0}} \right] + \\
 + 2{{\tilde \eta }^{\rm{T}}}{\dot {\tilde \eta}}  + 2\tilde x_{\delta 0}^{\rm{T}}{{\dot {\tilde x}}_{\delta 0}} = {\gamma _0}\left[  - \mu {{\tilde \xi }^{\rm{T}}}P\tilde \xi  - {{\tilde \xi }^{\rm{T}}}Q{Q^{\rm{T}}}\tilde \xi +\right.\hfill\\
 \hfill\left.+ 2{{\tilde \xi }^{\rm{T}}}P{{\hat \phi }^{\rm{T}}}\tilde \eta  + 2{{\tilde \xi }^{\rm{T}}}Q{\psi _d}h_\delta ^{\rm{T}}{\Phi _\delta }{{\tilde x}_{\delta 0}} \right] + \\
 + 2{{\tilde \eta }^{\rm{T}}}{\dot {\tilde \eta}}  + 2\tilde x_{\delta 0}^{\rm{T}}{{\dot {\tilde x}}_{\delta 0}} = {\gamma _0}\left[  - \mu {{\tilde \xi }^{\rm{T}}}P\tilde \xi  - {\textstyle{1 \over 2}}{{\tilde \xi }^{\rm{T}}}Q{Q^{\rm{T}}}\tilde \xi +\right.\hfill\\
\hfill \left. + 2{{\tilde \xi }^{\rm{T}}}P{{\hat \phi }^{\rm{T}}}\tilde \eta   - {\textstyle{1 \over 2}}{{\tilde \xi }^{\rm{T}}}Q{Q^{\rm{T}}}\tilde \xi  + 2{{\tilde \xi }^{\rm{T}}}Q{\psi _d}h_\delta ^{\rm{T}}{\Phi _\delta }{{\tilde x}_{\delta 0}} \pm \right.\\
 \left.\pm 2\tilde x_{\delta 0}^{\rm{T}}\Phi _\delta ^{\rm{T}}{h_\delta }\psi _d^{\rm{T}}{\psi _d}h_\delta ^{\rm{T}}{\Phi _\delta }{{\tilde x}_{\delta 0}} \pm 2{{\tilde \eta }^{\rm{T}}}\hat \phi {{\hat \phi }^{\rm{T}}}\tilde \eta  \right] +\hfill\\
 \hfill+ 2{{\tilde \eta }^{\rm{T}}}{\dot {\tilde \eta}}  + 2\tilde x_{\delta 0}^{\rm{T}}{{\dot {\tilde x}}_{\delta 0}}.
\end{gathered}
\end{equation}

Hereinafter the following change of notation is introduced $\phi \left( {y{\rm{,\;}}u{\rm{,\;}}\hat \delta } \right){\rm{:}} = \hat \phi $ for the sake of brevity.

Completing the square in \eqref{eqA7}, it is obtained:
\begin{equation}\label{eqA8}
\begin{gathered}
{{\dot V}_{\tilde \xi }} = {\gamma _0}\left[  - \mu {{\tilde \xi }^{\rm{T}}}P\tilde \xi  - {{\left( {{\textstyle{1 \over {\sqrt 2 }}}{{\tilde \xi }^{\rm{T}}}Q - \sqrt 2 {{\tilde \eta }^{\rm{T}}}\hat \phi } \right)}^2} -\right.\hfill\\\hfill
\left.- {{\left( {{\textstyle{1 \over {\sqrt 2 }}}{{\tilde \xi }^{\rm{T}}}Q - \sqrt 2 \tilde x_{\delta 0}^{\rm{T}}\Phi _\delta ^{\rm{T}}{h_\delta }\psi _d^{\rm{T}}} \right)}^2} +  \right.\\
 + \left. {2\tilde x_{\delta 0}^{\rm{T}}\Phi _\delta ^{\rm{T}}{h_\delta }\psi _d^{\rm{T}}{\psi _d}h_\delta ^{\rm{T}}{\Phi _\delta }{{\tilde x}_{\delta 0}} + 2{{\tilde \eta }^{\rm{T}}}\hat \phi {{\hat \phi }^{\rm{T}}}\tilde \eta } \right] +\hfill\\\hfill
 + 2{{\tilde \eta }^{\rm{T}}}{\dot {\tilde \eta}}  + 2\tilde x_{\delta 0}^{\rm{T}}{{\dot {\tilde x}}_{\delta 0}} \le \\
 \le {\gamma _0}\left[  - \mu {{\tilde \xi }^{\rm{T}}}P\tilde \xi  + 2\tilde x_{\delta 0}^{\rm{T}}\Phi _\delta ^{\rm{T}}{h_\delta }\psi _d^{\rm{T}}{\psi _d}h_\delta ^{\rm{T}}{\Phi _\delta }{{\tilde x}_{\delta 0}} + \right.\hfill\\\hfill
 \left. + 2{{\tilde \eta }^{\rm{T}}}\hat \phi {{\hat \phi }^{\rm{T}}}\tilde \eta  \right] + 2{{\tilde \eta }^{\rm{T}}}{\dot {\tilde \eta}}  + 2\tilde x_{\delta 0}^{\rm{T}}{{\dot {\tilde x}}_{\delta 0}}.
\end{gathered}
\end{equation}

If $\varphi \left( t \right) \in {\rm{FE}}$, then, following the results of Lemma 2, for all $t \ge {t_e}$ it holds that:
\begin{equation}\label{eqA9}
\begin{gathered}
\Delta \left( t \right) \ge {\Delta _{{\rm{min}}}} \ge \rho  > 0{\rm{,\;}}\left| {{{\cal M}_{{x_{\delta 0}}}}\left( t \right)} \right| \ge \underline {{{\cal M}_{{x_{\delta 0}}}}}  > 0{\rm{,}}
\end{gathered}
\end{equation}
then, according to the definition of the adaptive gains \eqref{eq43}, from \eqref{eqA8} it holds for all $t \ge {t_e}$ that:
\begin{equation}\label{eqA10}
\begin{gathered}
{{\dot V}_{\tilde \xi }} =  - \mu {\gamma _0}{{\tilde \xi }^{\rm{T}}}P\tilde \xi  + 2{\gamma _0}\tilde x_{\delta 0}^{\rm{T}}\Phi _\delta ^{\rm{T}}{h_\delta }\psi _d^{\rm{T}}{\psi _d}h_\delta ^{\rm{T}}{\Phi _\delta }{{\tilde x}_{\delta 0}} +\hfill\\\hfill
+ 2{\gamma _0}{{\tilde \eta }^{\rm{T}}}\hat \phi {{\hat \phi }^{\rm{T}}}\tilde \eta  - 2{{\tilde \eta }^{\rm{T}}}\left( {{\gamma _1} + {\gamma _0}{\lambda _{{\rm{max}}}}\left( {\hat \phi {{\hat \phi }^{\rm{T}}}} \right)} \right)\tilde \eta - \\
- 2\tilde x_{\delta 0}^{\rm{T}}{\gamma _{{x_{\delta 0}}}}{{\tilde x}_{\delta 0}} \le - \mu {\gamma _0}{{\tilde \xi }^{\rm{T}}}P\tilde \xi  - 2{\gamma _1}{{\tilde \eta }^{\rm{T}}}\tilde \eta - \hfill\\\hfill
- 2\tilde x_{\delta 0}^{\rm{T}}\left( {{\gamma _{{x_{\delta 0}}}} - c} \right){{\tilde x}_{\delta 0}} \le \\
\le- {\rm{min}}\left\{ {{\textstyle{{\mu {\gamma _0}{\lambda _{{\rm{min}}}}\left( P \right)} \over {{\lambda _{{\rm{max}}}}\left( P \right)}}}{\rm{, 2}}{\gamma _1}{\rm{,\;}}2{\gamma _{{x_{\delta 0}}}} \!-\! c} \right\}V{\rm{,\;}}
\end{gathered}
\end{equation}
from which the exponential convergence of errors $\tilde \xi \left( t \right){\rm{,\;}}\tilde \eta \left( t \right){\rm{,\;}}{\tilde x_{\delta 0}}\left( t \right)$ to zero is concluded regardless of the boundedness of $u\left( t \right)$ and $y\left( t \right)$.

The next aim is to prove the convergence of ${\tilde x_p}\left( t \right)$. Using \eqref{eqA10}, it is concluded that the multiplication ${T_I}\left( \theta  \right)\tilde \xi \left( t \right)$ converges exponentially to zero. Therefore, according to \eqref{eqA4}, to complete the proof it is necessary to prove that the virtual error $m\left( t \right) = {\tilde T_I}\left( t \right)\hat \xi \left( t \right)$ converges exponentially to zero too. For this purpose, we write a differential equation with respect to $m\left( t \right)$:
\begin{equation}\label{eqA11}
\begin{gathered}
\dot m\left( t \right) = {{\dot {\tilde T}}_I}\left( t \right)\hat \xi \left( t \right) + {{\tilde T}_I}\left( t \right){\dot {\hat \xi}} \left( t \right) = \hfill\\\hfill
= - {\gamma _{{T_I}}}\left( t \right){\cal M}_{{T_I}}^2\left( t \right){{\tilde T}_I}\left( t \right)\hat \xi \left( t \right) + {{\tilde T}_I}\left( t \right){\dot {\hat \xi}} \left( t \right) = \\\hfill
 =  - {\gamma _{{T_I}}}\left( t \right){\cal M}_{{T_I}}^2\left( t \right)m\left( t \right) + {{\tilde T}_I}\left( t \right){\dot {\hat \xi}} \left( t \right).
\end{gathered}
\end{equation}

Owing to ${{\cal M}_{{T_I}}}\left( t \right) \in \mathbb{R}{\rm{,\;}}m\left( t \right) \in {\mathbb{R}^3}$, we write the following equation after vectorization of the left- and right-hand sides of \eqref{eqA11} and application of $vec\left( {AB} \right) = \left( {{B^{\rm{T}}} \otimes I} \right)vec\left( A \right)$:
\begin{equation}\label{eqA12}
\begin{gathered}
\dot m\left( t \right) =  - {\gamma _{{T_I}}}\left( t \right){\cal M}_{{T_I}}^2\left( t \right)m\left( t \right) + \hfill\\\hfill
+ \left( {{{\dot {\hat \xi}^{\rm{T}}}}\left( t \right) \otimes {I_3}} \right)vec\left( {{{\tilde T}_I}\left( t \right)} \right) = \\
 =  - {\gamma _{{T_I}}}\left( t \right){\cal M}_{{T_I}}^2\left( t \right)m\left( t \right) + \left( {{{\dot {\hat \xi}^{\rm{T}}}}\left( t \right) \otimes {I_3}} \right){T_{Iv}}\left( t \right){\rm{,}}
\end{gathered}
\end{equation}
where $vec\left( {{{\tilde T}_I}\left( t \right)} \right) = {T_{Iv}}\left( t \right)$.

Again owing to ${\mathcal{M}_{{T_I}}}\left( t \right) \in \mathbb{R}$, the differential equation for ${T_{Iv}}\left( t \right)$  is written as:
\begin{equation}\label{eqA13}
\begin{gathered}
{\dot {\tilde T}_{Iv}}\left( t \right) =  - {\gamma _{{T_I}}}\left( t \right)\mathcal{M}_{{T_I}}^2\left( t \right){\tilde T_{Iv}}\left( t \right).
\end{gathered}
\end{equation}

A Lyapunov function candidate is introduced:
\begin{equation}\label{eqA14}
\begin{gathered}
{V_{{{\tilde x}_p}}}\left( t \right) = {\gamma _0}{m^{\rm{T}}}\left( t \right)m\left( t \right) + \tilde T_{Iv}^{\rm{T}}\left( t \right){\tilde T_{Iv}}\left( t \right).
\end{gathered}
\end{equation}

The derivative of \eqref{eqA14} is obtained as: 
\begin{equation}\label{eqA15}
\begin{gathered}
{{\dot V}_{{{\tilde x}_p}}} = {\gamma _0}\left( {{{\dot m}^{\rm{T}}}m + {m^{\rm{T}}}\dot m} \right) + 2\tilde T_{Iv}^{\rm{T}}{{\dot {\tilde T}}_{Iv}} = \hfill\\\hfill
 = {\gamma _0}\left(  - {\gamma _{{T_I}}}{\cal M}_{{T_I}}^2{m^{\rm{T}}}m + \tilde T_{Iv}^{\rm{T}}{{\left( {{{\dot {\hat \xi}^{\rm{T}}}} \otimes {I_3}} \right)}^{\rm{T}}}m -\right.\\
 \left.-\! {\gamma _{{T_I}}}{\cal M}_{{T_I}}^2{m^{\rm{T}}}m + {m^{\rm{T}}}\left( {{{\dot {\hat \xi}^{\rm{T}}}} \otimes {I_3}} \right){{\tilde T}_{Iv}} \right) \!+\! 2\tilde T_{Iv}^{\rm{T}}{{\dot {\tilde T}}_{Iv}} \!=\! \hfill\\\hfill
 = {\gamma _0}\left(  - 2{\gamma _{{T_I}}}{\cal M}_{{T_I}}^2{m^{\rm{T}}}m + 2\tilde T_{Iv}^{\rm{T}}{{\left( {{{\dot {\hat \xi}^{\rm{T}}}} \otimes {I_3}} \right)}^{\rm{T}}}\times\right.\hfill\\\hfill
 \left.\times\left( {{{\hat \xi }^{\rm{T}}} \otimes {I_3}} \right){{\tilde T}_{Iv}} \right) + 2\tilde T_{Iv}^{\rm{T}}{{\dot {\tilde T}}_{Iv}} = \\
 \le {\gamma _0}\left(  - 2{\gamma _{{T_I}}}{\cal M}_{{T_I}}^2{{\left\| m \right\|}^2} + 2{\lambda _{{\rm{max}}}}\left( {\dot {\hat \xi} {{\hat \xi }^{\rm{T}}} \otimes {I_3}} \right)\times\right.\hfill\\\hfill
 \left.\times {{\left\| {{{\tilde T}_{Iv}}} \right\|}^2} \right) - 2{\gamma _{{T_I}}}{\cal M}_{{T_I}}^2{\left\| {{{\tilde T}_{Iv}}} \right\|^2}
\end{gathered}
\end{equation}
where
\begin{displaymath}
\begin{gathered}
vec\left( {m\left( t \right)} \right) = vec\left( {{{\tilde T}_I}\left( t \right)\hat \xi \left( t \right)} \right){\rm{ = }}\left( {{{\hat \xi }^{\rm{T}}} \otimes {I_3}} \right){{\tilde T}_{Iv}}\left( t \right){\rm{,\;}}\\
{\left( {{{\dot {\hat \xi}^{\rm{T}}}} \otimes {I_3}} \right)^{\rm{T}}}\left( {{{\hat \xi }^{\rm{T}}} \otimes {I_3}} \right) = \left( {\dot {\hat \xi}  \otimes {I_3}} \right)\left( {{{\hat \xi }^{\rm{T}}} \otimes {I_3}} \right) = \dot {\hat \xi} {{\hat \xi }^{\rm{T}}} \otimes {I_3}.
\end{gathered}
\end{displaymath}

If $\varphi \left( t \right) \in {\rm{FE}}$, then, following the results of Lemma 2, for all $t \ge {t_e}$ the inequalities hold:
\begin{equation}\label{eqA16}
\begin{gathered}
\Delta \left( t \right) \ge {\Delta _{{\rm{min}}}} \ge \rho  > 0{\rm{,\;}}\left| {{{\cal M}_{{T_I}}}\left( t \right)} \right| \ge \underline {{{\cal M}_{{T_I}}}}  > 0{\rm{,}}
\end{gathered}
\end{equation}
and, therefore, according to the definition \eqref{eq43} of the adaptive gain ${\gamma _{{T_I}}}$, it follows from \eqref{eqA15} for all $t \ge {t_e}$ that:
\begin{equation}\label{eqA17}
\begin{gathered}
{{\dot V}_{{{\tilde x}_p}}} \le  - 2{\gamma _0}\left( {{\gamma _1} + {\gamma _0}{\lambda _{{\rm{max}}}}\left( {\dot {\hat \xi} {{\hat \xi }^{\rm{T}}} \otimes {I_3}} \right)} \right){\left\| m \right\|^2} +\hfill\\\hfill
+ 2{\gamma _0}{\lambda _{{\rm{max}}}}\left( {\dot {\hat \xi} {{\hat \xi }^{\rm{T}}} \otimes {I_3}} \right){\left\| {{{\tilde T}_{Iv}}} \right\|^2} - \\
 - 2\left( {{\gamma _1} + {\gamma _0}{\lambda _{{\rm{max}}}}\left( {\dot {\hat \xi} {{\hat \xi }^{\rm{T}}} \otimes {I_3}} \right)} \right){\left\| {{{\tilde T}_{Iv}}} \right\|^2} \le\hfill\\\hfill
 \le - 2{\gamma _0}{\gamma _1}{\left\| m \right\|^2} - 2{\gamma _1}{\left\| {{{\tilde T}_{Iv}}} \right\|^2} \le \\
 \le  - {\rm{min}}\left\{ {2{\gamma _0}{\gamma _1}{\rm{,\;}}2{\gamma _1}} \right\}{V_{{{\tilde x}_p}}}.\hfill
\end{gathered}
\end{equation}

From \eqref{eqA17} it is concluded that the multiplication \linebreak $m\left( t \right) = {\tilde T_I}\left( t \right)\hat \xi \left( t \right)$ and parametric error ${\tilde T_I}\left( t \right)$ converge exponentially to zero for all $t \ge {t_e}$, which, together with the exponential convergence to zero of both ${T_I}\left( \theta  \right)\tilde \xi \left( t \right)$ and $\tilde \eta \left( t \right){\rm{,\;}}{\tilde x_{\delta 0}}\left( t \right)$ for all $t \ge {t_e}$ proved in \eqref{eqA10}, results in exponential convergence of the augmented error ${\zeta _o}\left( t \right)$ for all $t \ge {t_e}$, which was to be proved.

\section{Proof of Theorem 2}
Proof is divided into two steps. The first one is to show that the error $\zeta \left( t \right)$ is bounded for all $t \in \left[ {{t_0}{\rm{,\;}}{t_e}} \right)$, the second one is to prove exponential convergence of $\zeta \left( t \right)$ for all $t \ge {t_e}$.

\textbf{Step 1}. The control law \eqref{eq18} is rewritten via the difference between \eqref{eq18} and \eqref{eq10}:
\begin{equation}\label{eqA18}
\begin{gathered}
u\left( t \right) = {{\hat \kappa }^{\rm{T}}}\left( t \right)\hat x\left( t \right) \pm {u^ * }\left( t \right) \pm {\kappa ^{\rm{T}}}\left( \theta  \right)\hat x\left( t \right) = \\
 = {{\hat \kappa }^{\rm{T}}}\left( t \right)\hat x\left( t \right) \pm {\kappa ^{\rm{T}}}\left( \theta  \right)x\left( t \right) \pm {\kappa ^{\rm{T}}}\left( \theta  \right)\hat x\left( t \right) = \\
 = {\kappa ^{\rm{T}}}\left( \theta  \right)x\left( t \right) + {{\tilde \kappa }^{\rm{T}}}\left( t \right)\hat x\left( t \right) + {\kappa ^{\rm{T}}}\left( \theta  \right)\tilde x\left( t \right) = \\
 = {\kappa ^{\rm{T}}}\left( \theta  \right)x\left( t \right) + {{\tilde \kappa }^{\rm{T}}}\left( t \right)\hat x\left( t \right) + {K_{x} }\left( \theta  \right){{\tilde x}_p}\left( t \right).
\end{gathered}
\end{equation}

The difference between the equations of the system \eqref{eq8} with the control law \eqref{eq10} and the same system \eqref{eq8} with \eqref{eqA18} is obtained as:
\begin{equation}\label{eqA19}
\begin{gathered}
{{\dot e}_{ref}}\left( t \right) = {\cal A}\left( \theta  \right)x\left( t \right) + {\cal B}\left( \theta  \right)\left( {\kappa ^{\rm{T}}}\left( \theta  \right)x\left( t \right) +\right.\hfill\\\hfill
\left.+ {{\tilde \kappa }^{\rm{T}}}\left( t \right)\hat x\left( t \right) + {K_{x} }\left( \theta  \right){{\tilde x}_p}\left( t \right) \right) + \\
 + {\cal D}\left( \theta  \right)\delta \left( t \right) - {\cal A}\left( \theta  \right){x^*}\left( t \right) -\hfill\\\hfill
 - {\cal B}\left( \theta  \right){\kappa ^{\rm{T}}}\left( t \right){x^*}\left( t \right) - {\cal D}\left( \theta  \right)\delta \left( t \right) = \\
 = \!{{\cal A}_{ref}}{e_{ref}}\left( t \right) \!+\! {\cal B}\left( \theta  \right)\left( {{{\tilde \kappa }^{\rm{T}}}\left( t \right)\hat x\left( t \right) + {K_{x} }\left( \theta  \right){{\tilde x}_p}\left( t \right)} \right).
\end{gathered}
\end{equation}

Considering conservative case, for all $t \in \left[ {{t_0}{\rm{,\;}}{t_e}} \right)$ it holds that $\Delta \left( t \right) < \rho  \Rightarrow {\gamma _\kappa }\left( t \right) = {\rm{0}}$ (it is assumed that $\Delta \left( t \right) = \linebreak = \rho  \Rightarrow t = {t_e}$), and, therefore, the identification procedure is not active $\dot {\tilde \kappa} \left( t \right) = 0$ over $\left[ {{t_0}{\rm{,\;}}{t_e}} \right)$, and it holds that:
\begin{equation}\label{eqA20}
\begin{gathered}
\tilde \kappa \left( t \right) = \tilde \kappa \left( {{t_0}} \right) < \infty .
\end{gathered}
\end{equation}

Moreover, using the results of Theorem 1, the state observation error ${\tilde x_p}\left( t \right)$ is bounded $\left\| {{{\tilde x}_p}\left( t \right)} \right\| \le \tilde x_p^{{\rm{max}}}$ for all $t \ge {t_0}$. It follows from the theorem of existence and uniqueness of solution of differential equations (Theorem 3.2 from \cite{b35}) that the error ${e_{ref}}\left( t \right)$ is bounded over $\left[ {{t_0}{\rm{,\;}}{t_e}} \right)$. From which, considering \eqref{eqA5}, we conclude that the augmented error $\zeta \left( t \right)$ is bounded over $\left[ {{t_0}{\rm{,\;}}{t_e}} \right)$.

\textbf{Step 2.} The aim is to show that the error $\zeta \left( t \right)$ converges exponentially to zero for all $t \ge {t_e}$. To this end, a Lyapunov function candidate is introduced:
\begin{equation}\label{eqA21}
\begin{gathered}
{V_{{e_{ref}}}} = {\gamma _0}e_{ref}^{\rm{T}}P{e_{ref}} + {\tilde \kappa ^{\rm{T}}}\tilde \kappa  + {\textstyle{{{c_0}} \over {{c_2}}}}{e^{ - {c_2}\left( {t - {t_e}} \right)}}{\rm{,}}
\end{gathered}
\end{equation}
where ${c_0} > 0{\rm{,\;}}{c_2} > 0$, and $P$ is a solution of the following set of equations under conditions $K = {k^2}{I_{n \times n}}$, $D = 0.5{k^2}{I_{n \times n}}{\rm{,\;}}k = 1{\rm{,\;}}B = {I_{n \times n}}$:
\begin{displaymath}
\begin{gathered}
{\cal A}_{ref}^{\rm{T}}P + P{{\cal A}_{ref}} =  - Q{Q^{\rm{T}}} - \mu P{\rm{,\;}}P{\cal B} = QK{\rm{,\;}}\\
{K^{\rm{T}}}K = D + {D^{\rm{T}}}{\rm{,}}
\end{gathered}
\end{displaymath}
which is equivalent to the Riccati equation with a bias term:
\begin{displaymath}
\begin{gathered}
{\cal A}_{ref}^{\rm{T}}P + P{{\cal A}_{ref}} + P{\cal B}{{\cal B}^{\rm{T}}}P + \mu P = {0_{\left( {n + 1} \right) \times \left( {n + 1} \right)}}.
\end{gathered}
\end{displaymath}

Owing to equations \eqref{eqA19}, \eqref{eq45}, \eqref{eq46}, the derivative of the quadratic form \eqref{eqA21} takes the form:
\begin{equation}\label{eqA22}
\begin{gathered}
{{\dot V}_{{e_{ref}}}} = {\gamma _0}e_{ref}^{\rm{T}}\left( {{\cal A}_{ref}^{\rm{T}}P + P{{\cal A}_{ref}}} \right){e_{ref}} + \hfill\\\hfill
 + {\gamma _0}\left[ e_{ref}^{\rm{T}}P{\cal B}{{\tilde \kappa }^{\rm{T}}}\hat x + {{\hat x}^{\rm{T}}}\tilde \kappa {{\cal B}^{\rm{T}}}P{e_{ref}} \!+\! e_{ref}^{\rm{T}}P{\cal B}{K_{x} }{{\tilde x}_p} \!+\right.\\
 \left. + \tilde x_p^{\rm{T}}K_x^{\rm{T}}{{\cal B}^{\rm{T}}}P{e_{ref}} \right] + 2{{\tilde \kappa }^{\rm{T}}}{\dot {\tilde \kappa}}  - {c_0}{e^{ - {c_2}\left( {t - {t_e}} \right)}} =\hfill\\\hfill
 = \!{\gamma _0}\!\left[  - \mu e_{ref}^{\rm{T}}P{e_{ref}} \!-\! e_{ref}^{\rm{T}}Q{Q^{\rm{T}}}{e_{ref}}  \!+\! 2e_{ref}^{\rm{T}}P{\cal B}{{\tilde \kappa }^{\rm{T}}}\hat x \!+\right.\\
 \left. + 2e_{ref}^{\rm{T}}P{\cal B}{K_{x} }{{\tilde x}_p} \right] + 2{{\tilde \kappa }^{\rm{T}}}{\dot {\tilde \kappa}}  - {c_0}{e^{ - {c_2}\left( {t - {t_e}} \right)}} =\hfill\\\hfill
 = \!{\gamma _0}\left[ { - \mu e_{ref}^{\rm{T}}P{e_{ref}} \!-\! {\textstyle{1 \over 2}}e_{ref}^{\rm{T}}Q{Q^{\rm{T}}}\!{e_{ref}} \!+\! 2e_{ref}^{\rm{T}}Q{{\tilde \kappa }^{\rm{T}}}\hat x \!- } \right.\\
\left.  - {\textstyle{1 \over 2}}e_{ref}^{\rm{T}}Q{Q^{\rm{T}}}{e_{ref}} \!+\! 2e_{ref}^{\rm{T}}Q{K_{x} }{{\tilde x}_p} \!\pm\! 2\tilde x_p^{\rm{T}}K_x^{\rm{T}}{K_{x} }{{\tilde x}_p} \!\pm \right.\hfill\\\hfill
\left. \pm 2{{\tilde \kappa }^{\rm{T}}}\hat x{{\hat x}^{\rm{T}}}\tilde \kappa \right] + 2{{\tilde \kappa }^{\rm{T}}}{\dot {\tilde \kappa}}  - {c_0}{e^{ - {c_2}\left( {t - {t_e}} \right)}}.
\end{gathered}
\end{equation}

Completing the square in \eqref{eqA22}, it is obtained:
\begin{equation}\label{eqA23}
\begin{gathered}
{{\dot V}_{{e_{ref}}}} \!=\! {\gamma _0}\left[  - \mu e_{ref}^{\rm{T}}P{e_{ref}} \!-\! {{\left( {{\textstyle{1 \over {\sqrt 2 }}}e_{ref}^{\rm{T}}Q \!-\! \sqrt 2 {{\tilde \kappa }^{\rm{T}}}\hat x} \right)}^2} \!- \right.\hfill\\\hfill
\left. - {{\left( {{\textstyle{1 \over {\sqrt 2 }}}e_{ref}^{\rm{T}}Q - \sqrt 2 \tilde x_p^{\rm{T}}K_x^{\rm{T}}} \right)}^2} + 2\tilde x_p^{\rm{T}}K_x^{\rm{T}}{K_{x} }{{\tilde x}_p}+ \right.\\
 \left.+ 2{{\tilde \kappa }^{\rm{T}}}\hat x{{\hat x}^{\rm{T}}}\tilde \kappa \right] + 2{{\tilde \kappa }^{\rm{T}}}{\dot {\tilde \kappa}}  - {c_0}{e^{ - {c_2}\left( {t - {t_e}} \right)}} \le \hfill\\\hfill
 \le {\gamma _0}\left[ { - \mu e_{ref}^{\rm{T}}P{e_{ref}} \!+\! 2\tilde x_p^{\rm{T}}K_x^{\rm{T}}{K_{x} }{{\tilde x}_p} \!+\! 2{{\tilde \kappa }^{\rm{T}}}\hat x{{\hat x}^{\rm{T}}}\tilde \kappa } \right] \!+\\\hfill
 + 2{{\tilde \kappa }^{\rm{T}}}{\dot {\tilde \kappa}}  - {c_0}{e^{ - {c_2}\left( {t - {t_e}} \right)}}.
\end{gathered}
\end{equation}

If $\varphi \left( t \right) \in {\rm{FE}}$, then, using results of Lemma 2, for all $t \ge {t_e}$ it holds that:
\begin{equation}\label{eqA24}
\begin{gathered}
\Delta \left( t \right) \ge {\Delta _{{\rm{min}}}} > 0{\rm{,\;}}\left| {{{\cal M}_\kappa }\left( t \right)} \right| \ge \underline {{{\cal M}_\kappa }}  > 0.
\end{gathered}
\end{equation}

According to results of Theorem 1, as the state observation error converges exponentially to zero regardless of the boundedness of $u\left( t \right)$ and $y\left( t \right)$, for all $t \ge {t_e}$ there exist some constants $c_1$ and $c_2$ such that:
\begin{equation}\label{eqA25}
\begin{gathered}
{\left\| {{{\tilde x}_p}\left( t \right)} \right\|^2} \le {c_1}{e^{ - {c_2}\left( {t - {t_e}} \right)}}.
\end{gathered}
\end{equation}

Then, according to the definition of ${\gamma _\kappa }\left( t \right)$ from \eqref{eq46}, we have for all $t \ge {t_e}$ from \eqref{eqA23} that:
\begin{equation}\label{eqA26}
\begin{gathered}
  {{\dot V}_{{e_{ref}}}} \leq  - \mu {\gamma _0}e_{ref}^{\text{T}}P{e_{ref}} + 2{\gamma _0}\tilde x_p^{\text{T}}K_x^{\text{T}}{K_{x}}{{\tilde x}_p} + \hfill\\\hfill
  +2{\gamma _0}{{\tilde \kappa }^{\text{T}}}\hat x{{\hat x}^{\text{T}}}\tilde \kappa  - 2{{\tilde \kappa }^{\text{T}}}\left( {{\gamma _1} + {\gamma _0}{\lambda _{{\text{max}}}}\left( {\hat x{{\hat x}^{\text{T}}}} \right)} \right)\tilde \kappa  -\\
  - {c_0}{e^{ - {c_2}\left( {t - {t_e}} \right)}} \leq  - \mu {\gamma _0}{\lambda _{{\text{min}}}}\left( P \right){\left\| {{e_{ref}}} \right\|^2} +\hfill\\\hfill
  +\! 2{\gamma _0}{\lambda _{{\text{max}}}}\left( {K_x^{\text{T}}{K_{x}}} \right){\left\| {{{\tilde x}_p}} \right\|^2} \!+\! 2{\gamma _0}{\lambda _{{\text{max}}}}\left( {\hat x{{\hat x}^{\text{T}}}} \right){\left\| {\tilde \kappa } \right\|^2} \!-  \\ 
   - 2\left( {{\gamma _1} + {\gamma _0}{\lambda _{{\text{max}}}}\left( {\hat x{{\hat x}^{\text{T}}}} \right)} \right){\left\| {\tilde \kappa } \right\|^2} - {c_0}{e^{ - {c_2}\left( {t - {t_e}} \right)}} =  \hfill\\ \hfill
   =  - \mu {\gamma _0}{\lambda _{{\text{min}}}}\left( P \right){\left\| {{e_{ref}}} \right\|^2} - 2{\gamma _1}{\left\| {\tilde \kappa } \right\|^2} -\\
   - \left( {{c_0} - 2{\gamma _0}{\lambda _{{\text{max}}}}\left( {K_x^{\text{T}}{K_{x}}} \right){c_1}} \right){e^{ - {c_2}\left( {t - {t_e}} \right)}} \leq \hfill\\ \hfill
    - {\text{min}}\!\left\{ {\tfrac{{\mu {\gamma _0}{\lambda _{{\text{min}}}}\left( P \right)}}{{{\lambda _{{\text{max}}}}\left( P \right)}}{\text{, 2}}{\gamma _1}{\text{, }}{c_0} \!-\! 2{\gamma _0}{\lambda _{{\text{max}}}}\left( {K_x^{\text{T}}{K_{x}}} \right){c_1}} \right\}\!V.
\end{gathered}
\end{equation}

The choice ${c_0} > 2{\gamma _0}{\lambda _{{\text{max}}}}\left( {K_x^{\text{T}}{K_{x}}} \right){c_1}$ allows one to ensure that ${c_0} - 2{\gamma _0}{\lambda _{{\text{max}}}}\left( {K_x^{\text{T}}{K_{x}}} \right){c_1} > 0$, from which the exponential convergence of errors ${e_{ref}}\left( t \right){\rm{,\;}}\tilde \kappa \left( t \right)$ to zero is concluded for all $t \ge {t_e}$, which, together with the exponential convergence of the observation error ${\tilde x_p}\left( t \right)$ proved in Theorem 1, results in exponential convergence of $\zeta \left( t \right)$ for all $t \ge {t_e}$.
}


\bf
\begin{IEEEbiographynophoto}{Anton Glushchenko}
received Software Engineering Degree from National University of Science and Technology "MISIS" (NUST "MISIS", Moscow, Russia) in 2008. In 2009 he gained Candidate of Sciences (Eng.) Degree from NUST "MISIS", in 2021 - Doctor of Sciences (Eng.) Degree from Voronezh State Technical University (Voronezh, Russia). Anton Glushchenko is currently the leading research scientist of laboratory 7 of V.A. Trapeznikov Institute of Control Sciences of Russian Academy of Sciences, Moscow, Russia. His research interests are mainly concentrated on exponentially stable adaptive control of linear time-invariant and time-varying plants under relaxed excitation conditions. The list of his works published includes more than 70 titles.
\end{IEEEbiographynophoto}

\vspace{-1cm}

\bf
\begin{IEEEbiographynophoto}{Konstantin Lastochkin}
received his bachelor’s degree in Electrical Engineering from National University of Science and Technology "MISIS" (NUST "MISIS", Moscow, Russia) in 2020, master's degree in Automation and control of technological processes - from NUST "MISIS" in 2022. Konstantin Lastochkin is currently a post-graduate student and a junior research scientist of laboratory 7 of V.A. Trapeznikov Institute of Control Sciences of Russian Academy of Sciences, Moscow, Russia. His present interests include adaptive and robust control, identification theory, nonlinear systems. The list of his works published includes 28 titles.
\end{IEEEbiographynophoto}

\vfill


\begin{thebibliography}{1}
\bibliographystyle{IEEEtran}
\bibitem{b1}
W. Leonhard {\it Control of electrical drives}, Springer Science \& Business Media, 2001.

\bibitem{b2}
R. Zhang, and C.Tong, ``Torsional vibration control of the main drive system of a rolling mill based on an extended state observer and linear quadratic control,'' \textit{Journal of Vibration and Control}, vol. 12, no. 3, pp. 313-327, 2006.

\bibitem{b3}
M. A. Valenzuela, J. M. Bentley, and R. D. Lorenz, ``Evaluation of torsional oscillations in paper machine sections,'' \textit{IEEE Transactions on Industry Applications}, vol. 41, no. 2, pp. 493-501, 2005.

\bibitem{b4}
R. Dhaouadi, K. Kubo, and M. Tobise, ``Two-degree-of-freedom robust speed controller for high-performance rolling mill drivers,'' \textit{IEEE Trans. Ind. Appl.}, vol. 29, no. 5, pp. 919–925, 1993.

\bibitem{b5}
G. Zhang, and J. Furusho, ``Speed control of two-inertia system by PI/PID control,'' \textit{IEEE Transactions on industrial electronics}, vol. 47, no. 3, pp. 603-609, 2000.

\bibitem{b6}
K. Szabat, and T. Orlowska-Kowalska, ``Vibration suppression in a two-mass drive system using PI speed controller and additional feedbacks -- Comparative study,'' \textit{IEEE Transactions on industrial electronics}, vol. 54, no. 2, pp. 1193-1206, 2007.

\bibitem{b7}
K. Sugiura, and Y. Hori, ``Vibration suppression in 2- and 3-mass system based on the feedback of imperfect derivative of the estimated torsional torque,'' \textit{IEEE Trans. Ind. Electron.}, vol. 43, no. 1, pp. 56–64, 1996.

\bibitem{b8}
K. Gierlotka, P. Zalesny, and M. Hyla, ``Additional feedback loops in the drives with elastic joints,'' in \textit{Proc. Int. Conf. EDPE}, Kosice, Slovakia, 1996, vol. 2, pp. 558–563.

\bibitem{b9}
J.M. Pacas, J. Armin, and T. Eutebach, ``Automatic identification and damping of torsional vibrations in high-dynamic-drives,'' in  \textit{Proc. ISIE}, Cholula-Puebla, Mexico, 2000, pp. 201–206.

\bibitem{b10}
K. Szabat, and T. Orlowska-Kowalska ``Comparative analysis of different PI/PID control structures for two-mass system,'' in \textit{Proc. 7th Int. Conf. Optim. Electr. and Electron. Equipment, OPTIM}, Brasov, Romania, 2004, pp. 97–102.

\bibitem{b11}
M.F.M. Yakub, A. Qadir, and B.A. Aminudin, ``Comparative study on control method for two-mass systems,'' \textit{International Journal on Advanced Science, Engineering and Information Technology}, vol. 2, no. 3, pp. 63-68, 2012.

\bibitem{b12}
K. Szabat, and T. Orlowska-Kowalska, ``Application of the Kalman filters to the high-performance drive system with elastic coupling,'' \textit{IEEE Transactions on Industrial Electronics}, vol. 59, no. 11, pp. 4226-4235, 2012.

\bibitem{b13}
P. Ioannou, and J. Sun, {\it Robust Adaptive Control}, N.Y.: Dover, 2013.

\bibitem{b14}
K.S. Narendra, and L.S. Valavani, ``Stable adaptive observers and controllers,'' \textit{Proceedings of the IEEE}, vol. 64, no.8, pp. 1198–1208, 1976.

\bibitem{b15}
G. Kreisselmeier, ``Adaptive observers with exponential rate of convergence,'' \textit{IEEE Transactions on Automatic Control}, vol. 22, no. 1, pp. 2-8, 1977.

\bibitem{b16}
J. K. Ji, and S. K. Sul, ``Kalman filter and LQ based speed controller for torsional vibration suppression in a 2-mass motor drive system,'' \textit{IEEE Transactions on industrial electronics}, vol. 42, no. 6, pp. 564-571, 1995.

\bibitem{b17}
K. Szabat, T. Orlowska-Kowalska, and K. P. Dyrcz, ``Extended Kalman filters in the control structure of two-mass drive system,'' \textit{Bulletin of the Polish Academy of Sciences: Technical Sciences}, pp. 315-325, 2006.

\bibitem{b18}
K. Szabat, and T. Orlowska-Kowalska, ``Performance improvement of industrial drives with mechanical elasticity using nonlinear adaptive Kalman filter,'' \textit{IEEE Transactions on industrial electronics}, vol. 55, no. 3, pp. 1075-1084, 2008.

\bibitem{b19}
K. Szabat, T. Orlowska-Kowalska, and M. Dybkowski, ``Indirect adaptive control of induction motor drive system with an elastic coupling,'' \textit{IEEE Transactions on Industrial Electronics}, vol. 56, no. 10, pp. 4038-4042, 2009.

\bibitem{b20}
Q. Zhang, ``Revisiting different adaptive observers through a unified formulation,'' in \textit{Proceedings of the 44th IEEE Conference on Decision and Control}, IEEE, Seville, Spain, 2005, pp. 3067-3072.

\bibitem{b21}
T. Orlowska-Kowalska, and K. Szabat, ``Neural-network application for mechanical variables estimation of a two-mass drive system,'' \textit{IEEE Transactions on Industrial Electronics}, vol. 54, no. 3, pp. 1352-1364, 2007.

\bibitem{b22}
T. Orlowska-Kowalska, and K. Szabat, ``Damping of torsional vibrations in two-mass system using adaptive sliding neuro-fuzzy approach,'' \textit{IEEE Transactions on Industrial Informatics}, vol. 4, no. 1, pp. 47-57, 2008.

\bibitem{b23}
T. Orlowska-Kowalska, and K. Szabat ``Control of the drive system with stiff and elastic couplings using adaptive neuro-fuzzy approach,'' \textit{IEEE Transactions on Industrial Electronics}, vol. 54, no. 1, pp. 228-240, 2007.

\bibitem{b24}
R. Ortega, ``Some remarks on adaptive neuro-fuzzy systems,'' in \textit{Proceedings of Tenth International Symposium on Intelligent Control}, IEEE, Monterey, California, USA, 1995, pp. 411-414.

\bibitem{b25}
A. Glushchenko, and K. Lastochkin, ``Parameter Estimation-Based Extended Observer for Linear Systems with Polynomial Overparametrization,'' \textit{arXiv preprint arXiv:2302.13705}, 2023, pp.1-6.

\bibitem{b26}
A. Glushchenko, and K. Lastochkin, ``Monotonous Parameter Estimation of One Class of Nonlinearly Parameterized Regressions without Overparameterization,'' \textit{arXiv preprint arXiv:2212.12184}, 2022, pp. 1-6.

\bibitem{b27}
R. Ortega, S. Aranovskiy, A. A. Pyrkin, A. Astolfi, and A. A. Bobtsov, ``New results on parameter estimation via dynamic regressor extension and mixing: Continuous and discrete-time cases,'' \textit{IEEE Transactions on Automatic Control}, 2020, vol. 66, no. 5, pp. 2265-2272.

\bibitem{b28}
V.O. Nikiforov, ``Observers of external deterministic disturbances. II. Objects with unknown parameters,'' \textit{Automation and Remote Control}, vol. 65, no.11, pp.1724–1732, 2004.

\bibitem{b29}
A.I. Glushchenko, K.A. Lastochkin, and V.A. Petrov, ``Exponentially stable adaptive control. Part I. Time-invariant plants,'' \textit{Automation and remote control}, vol. 83, no. 4, pp. 548-578, 2022.

\bibitem{b30}
P. J. Antsaklis, and A. N. Michel, {\it Linear systems}, New York: McGraw-Hill, 1997.

\bibitem{b31}
A. Glushchenko, and K. Lastochkin, ``Supplement to "Parameter Estimation-Based Extended Observer for Linear Systems with Polynomial Overparametrization,'' \textit{arXiv preprint arXiv:2302.13705}, pp.1–6, 2023, \url{https://arxiv.org/src/2302.13705v1/anc/CDC_PEBO_supp.pdf}.

\bibitem{b32}
A. Glushchenko, and K. Lastochkin, ``Exponentially Stable MRAC of MIMO Switched Systems with Matched Uncertainty and Completely Unknown Control Matrix,'' \textit{arXiv preprint arXiv:2208.03972}, 2022. pp. 1-6.

\bibitem{b33}
A. Glushchenko, V. Petrov, and K. Lastochkin, ``Regression filtration with resetting to provide exponential convergence of MRAC for plants with jump change of unknown parameters,'' \textit{IEEE Transactions on Automatic Control}, 2022, early access, pp.1-8.

\bibitem{b34}
G. Kreisselmeier, and G. Rietze-Augst, ``Richness and excitation on an interval-with application to continuous-time adaptive control,'' \textit{IEEE transactions on automatic control}, vol. 35, no. 2, pp. 165-171, 1990.

\bibitem{b35}
H. Khalil, {\it Nonlinear Systems}, 3rd ed., Upper Saddle River, NJ, USA: Prentice-Hall, 2002.

\end{thebibliography}
\end{document}